\documentclass[a4paper,11pt]{article}
\usepackage[utf8]{inputenc}
\usepackage{float, array, amscd, amsthm, amsmath, amssymb}

\usepackage[margin=1in,footskip=0.25in]{geometry}
\usepackage{hyperref}
\usepackage{color}
\usepackage{graphicx}
\usepackage{cite}
\usepackage[ruled,vlined]{algorithm2e}
\usepackage{longtable}
\usepackage{comment}
\newcommand{\cA}{\mathcal{A}}
\newcommand{\cO}{\mathcal{O}}

\newcommand{\cN}{\mathcal{N}}
\newcommand{\cE}{\mathcal{E}}
\newcommand{\cM}{\mathcal{M}}
\newcommand{\bZ}{\mathbb{Z}}
\newcommand{\bP}{\mathbb{P}}

\newcommand{\be}{\begin{equation}}
\newcommand{\ee}{\end{equation}}

\title{Heterotic Line Bundle Models on Generalized Complete Intersection Calabi Yau Manifolds}
\author{Magdalena Larfors\thanks{magdalena.larfors@physics.uu.se}, \qquad Davide Passaro\thanks{d.passaro@uva.nl}, \qquad Robin Schneider\thanks{robin.schneider@physics.uu.se}\\ \\
Department of Physics and Astronomy, Uppsala University\\ SE-751 20 Uppsala, Sweden\\
Institute of Physics, University of Amsterdam\\
Amsterdam, the Netherlands
}
\date{\today}

\begin{document}

\maketitle

\begin{abstract}
The systematic program of heterotic line bundle model building has resulted in a wealth of standard-like models (SLM) for particle physics. In this paper, we continue this work in the setting of generalised Complete Intersection Calabi Yau (gCICY) manifolds. Using the gCICYs constructed in  Ref.~\cite{Anderson:2015iia}, we identify two geometries that, when combined with line bundle sums, are directly suitable for heterotic GUT models. We then show that these gCICYs admit freely acting $\mathbb{Z}_2$ symmetry groups, and are thus amenable to Wilson line breaking of the GUT gauge group to that of the standard model. We proceed to a systematic scan over line bundle sums over these geometries, that result in 99 and 33 SLMs, respectively. For the first class of models, our results may be compared to line bundle models on homotopically equivalent Complete Intersection Calabi Yau manifolds. This shows that the number of realistic configurations is of the same order of magnitude. 
\end{abstract}

\newpage
\tableofcontents

\section{Introduction}

Calabi Yau (CY) manifolds are of particular importance for string theory. Compactifying string theory on CY manifolds produces lower dimensional effective field theories that preserve a certain fraction of supersymmetry. As such, these manifolds have, since the early days of string phenomenology, been the prime choice to construct string models of the four dimensional observable universe. CY manifolds also provide an arena for formal studies of string theory and have been instrumental in classifications of quantum field theories.

These diverse applications require explicit examples of the Calabi Yau geometry, and fortunately these exist in abundance. First, a CY $n$-fold may be constructed as an elliptic fibration over a base $(n-1)$-fold with certain regularity properties, see e.g. \cite{Morrison:1996pp}. Second, the CY manifold may be constructed as a hypersurface, or as a complete intersection of hypersurfaces, in a higher-dimensional K\"ahler ambient space. Taking the ambient space to be a toric variety results in the Kreuzer--Skarke list of CY manifolds \cite{Kreuzer:2000xy}. When the ambient space is instead a product of projective spaces, the resulting manifold is known as a Complete Intersection Calabi Yau (CICY). These have been completely classified in the case for 3- and 4-folds \cite{Candelas:1987kf,Gray:2013mja}. 

Notably, all CY manifolds constructed as (intersections of) hypersurfaces in ambient space are defined as the zero set of a (collection of) homogeneous polynomial equations. In the CICY case, this information may be captured in a configuration matrix, with positive integer entries. Half a decade ago, it was  realised \cite{Anderson:2015iia} that the CICY manifolds may be generalized to also include hypersurfaces defined by rational constraints, or, equivalently, configuration matrices that also have negative entries.  This led to new type of CY manifolds, dubbed gCICYs, and include examples that are topologically distinct from manifolds resulting from other CY constructions. Starting with the pioneering work of \cite{Anderson:2015iia}, the mathematical properties of these gCICYs have been charted by several authors  \cite{Berglund:2016yqo,Berglund:2016nvh,Candelas:2016fdy,Garbagnati:2017rtb,Jia:2018iza}. In contrast, the physical properties of string compactifications of gCICYs remain largely unexplored.\footnote{To the best knowledge of the authors, there is only one study where gCICYs have been explicitly used for string compactifications: in Ref.\cite{Anderson:2015yzz}, the authors show that gCICY 4-folds admit divisors that, when wrapped with M5 branes, lead to non-perturbative contributions to the superpotential in M theory compactifications.}
  
The heterotic string provides a  particular interesting setting for phenomenological studies of CY compactifications. Compactifying ten dimensional heterotic string theory on a CY 3-fold, dressed with a vector bundle satisfying the Hermitian-Yang-Mills equation, results in a four-dimensional, minimally supersymmetric, grand unified (GUT) model for particle physics \cite{Candelas:1985en}. If the CY manifold furthermore has non-trivial first fundamental group, the GUT gauge group may be broken to the  standard model gauge group via a Wilson line embedding. Adopting this strategy, the first heterotic CY vacua that match key features of the the minimally supersymmetric standard model were constructed  \cite{Braun:2005ux,Braun:2005bw,Bouchard:2005ag,Blumenhagen:2006ux}. More recently, taking the vector bundle to be a sum of line bundles has proven to be particularly useful, because of their ease of construction. Such configurations were first introduced in \cite{Anderson:2011ns,Anderson:2012yf} on CICYs and are, to date, the most successful construction of heterotic standard-like models (SLM) in terms of number of models found \cite{Anderson:2013xka,Constantin:2018xkj}.\footnote{The  term standard-like model (SLM) was coined in \cite{Deen:2020dlf}, and we adopt the same meaning as in that paper.} Furthermore, heterotic line bundle SLM have also been successfully constructed on hypersurfaces in toric varieties and elliptically fibered Calabi Yaus \cite{He:2013ofa,Braun:2017feb}. In this paper, we continue this program for heterotic line bundle model building by applying it to the gCICY manifolds.

{Before we go on, it is important to emphasise that the heterotic line bundle program, which we here apply to a novel set of geometries, is still far from producing fully realistic models of particle physics. Indeed, as emphasised in Refs.~ \cite{Anderson:2013xka,Constantin:2018xkj}, it is only by restricting attention to semi-topological properties of the compactification, that one can currently hope to make systematic studies. Particularly, as the name suggest, the  standard-like models that result from these constructions only reproduce some features of the standard model. To determine more detailed properties, such as masses and couplings,  would require knowledge of the CY metric, which is unavailable in analytic form for any compact CY. Furthermore, an important caveat of heterotic CY vacua is that the classical stability of these solutions may be compromised by non-perturbative effects such as world-sheet instantons \cite{Dine:1986zy, Dine:1987bq}. This is a subtle question as, while such contributions may be individually non-trivial, the sum of all the contributions from all world-sheet instantons may nevertheless vanish \cite{Distler:1986wm,Distler:1987ee,Silverstein:1995re,Beasley:2003fx}. Moreover, the compactifications we study are accompanied by geometric and bundle moduli \cite{Candelas:1990pi,MR0112154}, which must be fixed in any realistic model.  Another necessity is to break supersymmetry. These are highly non-trivial challenges, whose solution require ingredients that go beyond the perturbative setting we focus on. With the exception of a few remarks in the conclusion, we will not discuss this further.}

{Given the long list of highly non-trivial constraints that a fully realistic heterotic model must satisfy, it is clear that any given pair of CY manifold and line bundle is likely to fail to produce a standard model. The success of the heterotic line bundle program therefore relies on sifting through large numbers of pairs of CYs and bundles, and identify those that satisfy phenomenological criteria. In adherence to this program, in this paper we are thus not trying to construct a phenomenologically realistic vacuum for physics. Instead we start working with a new, possibly infinite, set of manifolds, in the hope that by constructing a large number of putative vacua and efficiently filtering out those that satisfy necessary physical constraints we may eventually narrow down on realistic models. Furthermore, as we will describe in detail below, our study contributes to determining which of the  different CY presentations are most effective for model building. Here, the number of heterotic SLMs produced by a CY presentation serves as a benchmark.
}

{With this precaution, there are several reasons why it is of interest to study of heterotic line bundle models on gCICY manifolds.} First, as mentioned above, the gCICY construction may result in manifolds that are topologically distinct from other CY constructions, and one should establish if these new manifolds, in combination with suitable vector bundles, satisfy the topological constraints required for heterotic SLM. While this may be expected, due to the similarity between gCICY and CICY manifolds, it requires overcoming several technical challenges, as we will soon explain. Having proven that this is the case, it is relevant to compare how successful heterotic line bundle model building is on gCICY manifolds as compared to other CY constructions. In particular, it is of interest to compare, for gCICY manifolds that may alternatively be described as e.g. CICY manifolds, if the same number of heterotic SLM is obtained in both constructions. Last, but not least, this study is motivated by the development of computational tools that will likely be of relevance for further geometrical studies of gCICY manifolds.

However, before we can  build 
heterotic models on gCICY manifolds, we must tackle a few technical challenges. The first, and most severe, obstacle to gCICY model building is the determination of the Kähler cone, which is, in particular, required to prove the stability of the vector bundle. In other CY constructions, this is under control for Kähler favourable manifolds, so that the $h^{(1,1)}(X)$ Kähler forms arise as restrictions of ambient space forms. We will conform with this restriction, which holds in a rather limited fraction of the many classes of gCICY manifolds constructed in Ref.~\cite{Anderson:2015iia}. Second, manifolds with low $h^{(1,1)}$ are expected to result in few heterotic line bundle models \cite{Anderson:2013xka,Constantin:2018xkj}. In fact, as we will expand on in section \ref{sec:lbmodels}, the vector bundle stability constraints can generically not be solved for $h^{(1,1)}<4$. Finally, even assuming Kähler favourability, delineating the Kähler cones becomes increasingly hard with growing codimension, and when  the rational constraints defining the gCICY have denominators that are quadratic or higher. Thus, it is opportune to restrict to gCICYs of low codimension and where the configuration matrix entries are larger or equal to -1.

With this motivation, in this paper, we select the two Kähler favourable gCICY manifolds with $h^{(1,1)}=5$ from Ref.~\cite{Anderson:2015iia} and systematically construct heterotic line bundle models on them. {One of these spaces, which we denote $X_1$, may also be realised as a  CICY, whereas the other, $X_2$ is topologically distinct from all CICY realisations. More to the point, $X_2$ is topologically distinct from all CY manifolds that have hitherto been used for heterotic SLM constructions.} We explicitly determine the topological properties of these gCICY, discuss their K\"ahler and Mori cones, and develop the machinery needed to determine the associated line bundle cohomologies. 

Finally, we prove that Wilson line symmetry breaking may be realised in the resulting models. Being defined by rational equations, it is in fact non-trivial to determine whether a gCICY manifold is simply connected; tools such as the Lefschetz hyperplane theorem do not apply in this situation \cite{Anderson:2015iia,Berglund:2016yqo}. While we will not resolve this matter, and therefor remain agnostic about the full first fundamental group of these gCICYs, we will prove that  both manifolds admit a freely acting $\mathbb{Z}_2$-symmetry. This leads to two smooth quotient manifolds with non trivial fundamental group that contains a $\mathbb{Z}_2$ component, and which thus allow Wilson line breaking of a GUT $SU(5)$ group to $SU(3)\times SU(2) \times U(1)$. We then proceed to scan over line bundle model configurations with increasing charges, finding in total about 130 heterotic standard-like models. 
 
 The outline of this paper is as follows. In section \ref{sec:lbmodels} we briefly recap the construction of heterotic line bundle models. Section \ref{sec:setting} is split into three parts, first a brief overview of the two geometries \ref{sec:twogcicys}, second a discussion of their K\"ahler cones \ref{sec:kaehler} and third the construction of the quotient manifolds \ref{sec:freely}. We proceed in section \ref{sec:results} with the results of the systematic scan and conclude in section \ref{sec:outlook}. The results of the scan, and technical discussions regarding topological invariants and Kähler cones, are presented in appendices.

\section{Heterotic Line Bundle Models}
\label{sec:lbmodels}

In this section we will briefly review the construction of heterotic standard-like models (SLM) in the context of line bundle models, a topic that was pioneered and discussed in great detail in Ref.~\cite{Anderson:2011ns,Anderson:2012yf,Anderson:2013xka}. Our aim is to construct vector bundles satisfying a set of  consistency and phenomenological constraints that are required for SLMs. Given a smooth Calabi Yau manifold with non trivial fundamental group, we define a vector bundle as a sum of five line bundles
\begin{align}
	\label{eq: V}
	V = \bigoplus_{a=1}^5 L_a = \bigoplus_{a=1}^5  \mathcal{O}_X(q^1_a , ..., q^{r}_a)
\end{align}
where for K\"ahler favourable CY manifolds $r = h^{(1,1)}(X)$. 
First, we impose the condition $c_1^i(V) = \sum_{a=1}^{5} q^i_a = 0$ such that $c_1(V)=c_1^i(V)J_i$ vanishes and $V$ has structure group $S(U(1)^5)$. 
Embedding the structure group into one of the $E_8$-factors of the heterotic string, we reach the intermediate GUT group 
\begin{align}
\label{eq:gauge}
    E_8 \supset SU(5) \times S(U(1)^5) \cong SU(5) \times U(1)^4.
\end{align}
We then require anomaly cancellation which can be derived from the integrability constraints of the 10D Bianchi identity; together with the Bogomolov bound for slope zero stable bundles this translates to the following condition on the second Chern class
\begin{align}
\label{eq:bianchi}
    &0 < c_2(V) \leq c_2 (X) \\
    &\text{with:} \qquad  c_{2\, i}(V) = - \frac{1}{2} d_{ijk} \sum_{a=1}^5 q^j_a q^k_a 
\end{align}
where $d_{ijk}$ are the triple intersection numbers of the underlying Calabi Yau 3-fold, and we have expanded $c_2$ with respect to a basis of four-forms. {The constraint \eqref{eq:bianchi} furthermore requires, unless the right hand inequality is saturated, that the resulting model contains space-filling NS5-branes, which wrap effective curves in the internal geometry \cite{Anderson:2013qca,Anderson:2012yf}. The world-volume theories of these branes will contribute hidden sector gauge fields to the four-dimensional models, as well as scalars given by the NS5 moduli. This is discussed in some detail in \cite{Lukas:1998hk,Braun:2005bw}. We will not discriminate between models that require NS5-branes, versus those who do not, in order to conform with the standard practice of the heterotic line bundle program.  }

Next, the Donaldson-Uhlenbeck-Yau theorem \cite{donaldson,uhlenbeck-yau} states that supersymmetry in the lower dimensional theory is preserved if and only if the vector bundle $V$ is polystable and has slope zero. We require this to hold at a locus in the K\"ahler cone, which is compatible with the supergravity approximation. For sums of line bundles it translates to the following condition on the Kähler moduli:
\begin{align}
\label{eq:slope}
    \exists t^1, t^2, ... , t^{h^{(1,1)}} > 1 : \mu(L_a) \stackrel{!}{=} 0, \forall a \in \{ 1,...,5 \}
\end{align}
where $t^i$ are the K{\"a}hler moduli and the slope of a line bundle is computed by
\begin{align}
\mu(L_a) = d_{ijk} q^i_a t^j t^k.
\end{align}
Note that the slope-zero condition \eqref{eq:slope} is very constraining for low $h^{(1,1)}$, and may require a significant tuning of line bundle charges $q^i_a$. With $h^{(1,1)}>4$, on the other hand, \eqref{eq:slope} can be solved for generic values of the charges.

\begin{table}[t]
    \centering
    \begin{tabular}{|c|c|c|}
    \hline
    multiplet & l.b. cohomologies counting multiplicities & contained in\\
    \hline
    10     &  $h^1(X, L_a)$ & $V$\\
    $\Bar{10}$     & $h^1(X, L_a^*)$& $V^*$\\
    5     &  $h^1(X, L_a \otimes L_b)$ & $V \wedge V$\\
    $\Bar{5}$     & $h^1(X, L_a^* \otimes L_b^*)$& $V^* \wedge V^*$\\
    1     &  $h^1(X, L_a \otimes L_b^*)$ and $h^1(X, L_a^* \otimes L_b)$& $V \otimes V^*$\\
    \hline
    \end{tabular}
    \caption{Relevant particle content of the SU(5) GUT theory and their associated line bundle cohomology counting the multiplicity.}
    \label{tab:content}
\end{table}
% Particle content
The relevant particle content of the $SU(5)$ GUT theory and its multiplicities are given in Table \ref{tab:content}. To further break the GUT group to the standard model one uses the fundamental group $\Gamma$ of the quotient manifold as a Wilson line, as explained e.g. in Ref.~\cite{Green:1987mn}. The number of multiplets after Wilson line breaking is then given by the invariant part of $H^1(X, V) \otimes R_W$, where $R_W$ is some Wilson line representation. A consequence of this is that for each $L_a \in V$ 
\begin{align}
    \chi(L_a) \mod |\Gamma| &= 0 \\
    \text{where } \chi(L_a) &= d_{ijk} \left( \frac{1}{6} q_a^i q_a^j q_a^k + \frac{1}{12} q_a^i c_2(X)^{jk} \right)
\end{align}
must hold true to build an equivariant structure descending to the downstairs manifold.
The particle content of the downstairs theory is then given by 
\begin{align}
    \# \text{fermion generations} = - \frac{\chi(V)}{|\Gamma|}.
\end{align}
Thus, in order to find exactly three fermion generations with no anti-families we require the physical constraint
\begin{align}
\label{eq:fermion}
h^\bullet(V) = (0,3 |\Gamma|,0,0).
\end{align}
Furthermore we want to have at least one pair of Higgs doublets, and no Higgs triplets which translates to the following conditions
\begin{align}
\label{eq:higgs}
    \text{doublet: } h^2(\wedge^2 V) > 0 , \qquad \text{triplet: } \text{ind}(L_a \otimes L_b) \leq 0 \; \forall a,b.
\end{align}
% U(1) Bosons and mass
 Clearly \eqref{eq:fermion}-\eqref{eq:higgs} shows that the compactification matches crucial properties of the standard model. More detailed phenomenological properties cannot be determined in this systematic manner, and are therefor not accounted for here.\footnote{There are also properties that can be determined, but that may be reconciled in different ways with the  standard model. For example, the downstairs spectrum has the gauge symmetry of the standard model plus the additional $U(1)$-factors coming from the Abelian split-locus \eqref{eq:gauge}. This introduces additional vector bosons to the theory. They gain masses by the Green-Schwarz mechanism \cite{Anderson:2011ns} with mass matrix
$M_{ab} = - \partial_i \partial_j \ln(\kappa) c_1^i(L_a) c_1^j(L_b)$,
where $\kappa$ is the Kähler potential. Thus the number of additional massless $U(1)$-charges is given by $4-\text{rank}(V)$.}

\section{gCICY Geometry}
\label{sec:setting}

In this section we will introduce the gCICY manifolds that are in focus in this paper. We will not review all details regarding the general construction of gCICYs, but instead discuss the geometric constraints that pertain to heterotic model building, with an emphasis on two explicit example manifolds. For more thorough discussions, the reader  is referred to Ref.~\cite{Anderson:2015iia}and the second author's master thesis \cite{Passaro:2020}. The mathematical properties of gCICY are further explored in   \cite{Berglund:2016yqo,Berglund:2016nvh,Candelas:2016fdy,Garbagnati:2017rtb,Jia:2018iza}.

Just as ordinary CICY manifolds, gCICYs are constructed as subvarieties in products of projective spaces ${\cal A}={\mathbb P}^{n_1} \times {\mathbb P}^{n_2} ... \times {\mathbb P}^{n_s}$. We will refer to $\cal{A}$ as the ambient space and let $X$ denote the gCICY subvariety. $X$ is a complete intersection, {\it i.e.~}the common zero locus of a number of homogeneous equations. Similar to a CICY, the full information of a gCICY is encoded in a configuration matrix \cite{Anderson:2015iia}:
\begin{align} \label{eq-config}
  X\in\left[ \begin{array}{c||ccc|ccc}
	  n_{1} & a^{1}_{1} & \dots & a^{1}_{K} & b^{1}_{1} & \dots & b^{1}_{L}\\
	  \vdots & \vdots & \ddots & \vdots & \vdots & \ddots &  \vdots \\
	  n_{s} & a^{s}_{1} & \dots & a^{s}_{K} & b^{s}_{1} & \dots & b^{s}_{L}
  \end{array}\right]^{h^{(1,1)},h^{(2,1)}}_{\chi}
\end{align}    
Here $n_i$ denote the degrees of the $i-$th projective component of the ambient space. The remaining integers $a_j^i$,  $b_j^i$ specify the homogeneous equations:  $a_j^i$ is the degree of the $j-$th polynomial constraint in the $i-$th ambient space variables, and $b_j^i$ are the corresponding degrees of rational constraints. As a consequence, $a_j^i$ are non-negative, whereas $b_j^i$ may  take any integer value. Thus, the first $K$ equations define a submanifold $\cM$ in which we embed the gCICY by the $L$ rational constraints. We say that the gCICY is of codimension $(K,L)$. We will return to the discussion of the topological invariants $h^{(1,1)},h^{(2,1)},\chi $ below. 

For the gCICY to be smooth and well-defined, $X$ must avoid the poles that the rational constraints would introduce on $\cA$. We will elaborate on this subtlety below. In short, what we will see is that, after first imposing the $K$ polynomial constraints, resulting in $\cM$, we must successively impose the $L$ rational constraints in the order specified by \eqref{eq-config}. This gives a succession of manifolds ${\cal M} \supset {\cal M}_1 \supset {\cal M}_2 \supset ...\supset {\cal M}_{L-1} \supset X$ where each manifold ${\cal M}_r$ avoids the poles of the constraint that defines ${\cal M}_{r+1}$. Note that this is a key difference with respect to CICY manifolds: polynomial constraints do not introduce poles and may therefore be applied in any order without jeopardizing the smoothness of the manifold.

In the remainder of this section, we will construct and analyze  two codimension (1,1) gCICY threefolds that we will use for the heterotic line bundle models discussed in section \ref{sec:lbmodels}. We will prove that these gCICYs are smooth, compute their topological properties, discuss their K\"ahler cones, and finally show that they admit freely acting $\mathbb{Z}_2$ symmetries. This puts us in a position to discuss the mathematical constraints imposed in heterotic line bundle models on these manifolds, which we will do in section \ref{sec:results}.

\subsection{Two gCICYs}
\label{sec:twogcicys}

In the rest of this paper we will be particularly interested in two gCICYs
\begin{align}\
\label{eq:gCICYs}
      X_{1}\in\left[ \begin{array}{c||c|c}
		1 & 1 & 1 \\
		1 & 1 & 1 \\
		1 & 1 & 1 \\
		1 & 1 & 1 \\
		1 & 3 & -1 
	\end{array}\right]_{-80}^{5,45} \qquad \text{and} \qquad 
    X_{2}\in\left[ \begin{array}{c||c|c}
		1 & 1 & 1 \\
		1 & 1 & 1 \\
		1 & 1 & 1 \\
		1 & 0 & 2 \\
		1 & 3 & -1 
	\end{array}\right]_{-48}^{5,29}.
\end{align}
These are smooth manifolds that were first constructed in Ref.~\cite{Anderson:2015iia}. $X_1$ also has a CICY realisation, while $X_2$ is topologically distinct from any CY manifold that has previously been used for heterotic line bundle model building. As mentioned in the Introduction, they are particularly  promising for model building  due to their relatively large $h^{(1,1)}=5$. 

\subsubsection*{Construction and Smoothness}
\label{sec:smoothness}

In this section, we will recapitulate the construction of the gCICY threefolds $X_1$ and $X_2$, and show that they are smooth manifolds. We include this detailed discussion for completeness, although the main results are already given in Ref.~\cite{Anderson:2015iia}, and since we will perform similar analyses in section \ref{sec:freely}. For brevity,  we will focus on $X_1$. The analysis for $X_2$ is completely analogous, and contained in  Ref.~\cite{Passaro:2020}.

As for all codimension (1,1) gCICYs, the construction of $X_1$ involves two steps: first, the manifold $\cM$ is the  submanifold of $\mathcal{A}$ defined by $p=0$, where the polynomial $p$ is chosen as a section of $\cO_\cA(1,1,1,1,3)$. Explicitly, if we let $y_{i,0},y_{i,1}, i=0,1,2,3,4$ be the homogeneous coordinates of the projective spaces, we read off from the configuration matrix \eqref{eq:gCICYs} that 
\begin{align}
  p=\sum_{i=0}^{3}y_{4,0}^{3-i}y_{4,1}^{i}d_{i},
  \label{eq:polyconstr1}
\end{align}
where the $d_{i}$ are generic linear polynomials in the homogeneous coordinates of the first four complex projective spaces.

Second, the gCICY is defined by a rational constraint $q=0$. We construct this as a linear combination of all linearly independent rational sections on $\cM$.  Utilizing computational tools that will be introduced below (in particular the Koszul sequence \eqref{seq:koszul}) we have
\begin{align}
    h^{0} \left( \cM_1, \cO_{\cM_1} ( 1,1,1,1,-1 ) \right) = 3 
    \; .
\end{align}
Hence, the submanifold $\cM$ admits 3 rational sections.  
To write out these as rational constraints on $\cA$, we first select a denominator. According to  \eqref{eq:gCICYs}, this should be linear, and we can take:
\begin{align*}
  \Delta_{1}=y_{4,0},\ \Delta_{2}=y_{4,0}-y_{4,1},\ \Delta_{3}=y_{4,0}+y_{4,1}.
\end{align*}
Focusing on the first denominator, $ \Delta_{1}=y_{4,0}$, we may then construct rational constraints on $\cA$ that are without poles on $\cM$, if the nominator $N_1$ of the rational function also vanishes whenever $\Delta_{1} = p =0$. Consider the following polynomial division: 
\be
  p= \Delta_{1}m_{1}+y_{4,1}^{3}N_{1} \; ,
\ee
where $m_1, N_1$ are polynomials to be determined. When $\Delta_{1} = p =0$ one has $y_{4,1}^{3}N_{1}=y_{4,1}^{3}d_{3}=0$.
Since the homogeneous coordinates of a $\mathbb{P}^1$ cannot vanish simultaneously, we must then have $d_{3}=0$ on $\cM$, and can use the linear polynomial $d_3$  for the construction of the numerator $N_1$. Analysing the other denominators, and the polynomial form of the $m_i$, we eventually find 
\begin{gather}
  N_{1}=d_{3},\ N_{2}=\sum_{i=0}^{3}d_{i},\ N_{3}=\sum_{i=0}^{3}\left( -1 \right)^{i+1}d_{i}\\
  m_{1}=y_{4,0}^2d_0+y_{4,0}y_{4,1}d_1+y_{4,1}^2d_{2}\\
  m_{2}=(y_{4,0}^2+y_{4,0}y_{4,1}+y_{4,1}^2)d_{0}+(y_{4,0}+y_{4,1})y_{4,1}d_{1}+y_{4,1}^2d_{2}\\
  m_{3}=(y_{4,0}^2-y_{4,0}y_{4,1}+y_{4,1}^2)d_{0}+(y_{4,0}-y_{4,1})y_{4,1}d_{1}+y_{4,1}^2d_{2}
\end{gather}
and may write the three rational sections for $\cM$ as
\be
\label{eq:rsection}
s_i = \frac{N_i}{\Delta_i} \; .
\ee

Finally, we write the rational section $q$ that defines $X_1$ as a submanifold in $\cM$  as a linear combination of the $s_i$.  In order to prove the smoothness of $X_1$, we need to determine the exact combination locally. Defining the four regions
\begin{align}
  \begin{array}{cccc}
  R_{0}: & \Delta_{1}\neq 0 & \Delta_{2}\neq 0 & \Delta_{3}\neq 0 \\
  R_{1}: & \Delta_{1}= 0 & \Delta_{2}\neq 0 & \Delta_{3}\neq 0 \\
  R_{2}: & \Delta_{1}\neq 0 & \Delta_{2} =0 & \Delta_{3}\neq 0 \\
  R_{3}: & \Delta_{1}\neq 0 & \Delta_{2}\neq 0 & \Delta_{3}= 0 .
\end{array} \label{eq:regiondivision}
\end{align}
where we note that at most one of the denominators $\Delta_i$ may vanish simultaneously, we may write for a generic rational section, 
\begin{align}
	R_{0}:&\quad   q_{0}= \alpha_{1} \frac{N_{1}}{\Delta_{1}}+\alpha_{2}\frac{N_{2}}{\Delta_{2}}+\alpha_{3}\frac{N_{3}}{\Delta_{3}}\label{eq:ratsec1R0}\\
	R_{i}:&\quad q_{i}=q_{0}-\alpha_{i}\frac{N_{i}}{\Delta_{i}}-\alpha_{i}\frac{m_{i}}{y_{4,1}^{3}}\label{eq:ratsec1Ri},
\end{align}
where $i=1,2,3$ and $\alpha_{i}$ are arbitrary constants.

We are now in the position to discuss the smoothness of $X_1$. This is more laborious than in other construction of CY manifolds, because of the rational constraints. To start, recall that $\mathcal{M}$ is defined by a generic polynomial constraint in a compact complex manifold. Consequently, by Bertini's theorem, $\cM$ is smooth. For $X_1$, however, we just showed that it is necessary to tune the rational constraint $q$ so that its poles are avoided on $\cM$, and thus the constraint is no longer generic. What does hold, however, is that $q$ is polynomial on $\cM$. Explicitly, we may use the defining relation $p=0$  for $\cM$ to rewrite all rational function as polynomials patchwise,  viz.
\begin{align*}
   \frac{N_{1}}{\Delta_{1}}\to \frac{m_{1}}{y_{4,1}^{3}}
\end{align*}
so that $s_1=0$ simply imposes the polynomial constraint $m_1=0$ in the patch $R_1: y_{4,1}\neq 0$.  Thus, in each region $R_{0,1,2,3}$ we have polynomial representations of all constraints defining $X_1$. We will append to this collection of polynomials a set of ``Lagrange constraints", with variables $z_i$
\begin{equation}
  L_{1}:=z_{1}\Delta_{1}-1,\quad L_{2}:=z_{2}\Delta_{2}-1,\quad L_{3}:=z_{3}\Delta_{3}-1.
\end{equation}
Note that $L_i$ can only be zero when the denominators $\Delta_i$ are non-zero, thus we may use the $L_i$ to encode the region $R_{0,1,2,3}$ that we wish to analyse. 
Using the set of polynomials obtained in this way we may determine the smoothness of $X_{1}$ with standard Gr\"obner basis techniques (see e.g. \cite{cox2013ideals}).
In particular, using a computer algebra system, one can define the ideals generated by $\left\{ p,q_{\alpha},dp\wedge dq_{\alpha},\rho_{\alpha},\phi_{U_{i}} \right\}$, where $p$ is the polynomial section, $q_{\alpha}$ is the numerator of the rational section in the region $R_{\alpha}$.
The polynomial (or set of polynomials) $\rho_{\alpha}$ localizes the computation on a particular region $R_{\alpha}$. $\rho_{\alpha}$ is the set $\left\{L_{1},L_{2},L_{3}  \right\}$ for $\alpha=0$, as these conditions constrain the denominators to be non-zero, and  $\rho_{\alpha}$ is $\Delta_{\alpha}$ for $\alpha=1,2,3$. Lastly, $\phi_{U_{i}}$ localizes the computation further on a coordinate chart.

By checking that each of the varieties, defined by the ideals above, are empty, one can assert the smoothness of $X_{1}$.
We have performed this check using Sage \cite{sagemath} as an interface to Singular \cite{DGPS}. For Mathematica users the STRINGVACUA interface comes with similar functionality \cite{Gray:2008zs}.

\subsection{Topological Quantities}
\label{sec:sequences}

As discussed in section \ref{sec:lbmodels}, topological quantities are of prime importance for heterotic model building. In particular, we will need the Chern classes and cohomology group dimensions of the manifold and vector bundle, and the triple intersection numbers of the manifold. In this section and appendix \ref{ap:topology}, we will outline how we compute these properties, referring the interested reader to Ref.~ \cite{Anderson:2015iia} for more details. Since these computations quickly become too involved to perform manually, we proceed using a modified version of the pyCICY package \cite{CICYtoolkit} developed in the context of \cite{Larfors:2019sie}. The relevant scripts can be found in the repository \cite{gCICYtoolkit}.

Just as for CICY manifolds, the topological invariants of $X$ are encoded in the configuration matrix \eqref{eq-config}. Moreover, since both CICYs and gCICYs are algebraic varieties, we may apply the same algebraic geometry tools to determine their topology. The rationale behind these tools is to relate the quantity of interest on $X$ to quantities that may be readily determined on the ambient space $\cA$ (using the Euler sequence, the Bott-Borel-Weil theorem and the K\"unneth formula). The novelty in the gCICY case is that we now need to apply the tools repeatedly for the  succession of manifolds $\cA \supset {\cal M} \supset {\cal M}_1 \supset {\cal M}_2...\supset {\cal M}_{L-1} \supset X$, thus computing the topological properties of $X$ in a stepwise manner. For codimension (1,1) gCICYs, there is thus one additional computational layer to consider: $\cA \supset {\cal M} \supset X$. 

In brief, we will need the following tools (see \cite{griffithsharris,hartshorne}  for their derivation and \cite{Anderson:2008ex,Hubsch:1992nu} for discussions in the setting of string compactifications): 
\begin{itemize} 
\item The Euler sequence, for $\cA$ a $k$-fold product of $n_i$-dimensional project spaces, reads
\be \label{eq:euler}
0 \to \cO^{\oplus k}_\cA \to \cO^{\oplus n_1 +1}_\cA (1,0,...,0) \oplus ... \oplus   \cO^{\oplus n_k +1}_\cA (0,0,...,1) \to T\cA \to 0
\ee
We will use this sequence, and tensor products of it, to determine the cohomologies of the ambient space $\cA$, and vector bundles over $\cA$.
\item The adjunction formula shows how the tangent bundle  of $\cA$, locally splits into the tangent bundle of an algebraic submanifold $\cM$ and its normal bundle $\cE_1$:
\begin{equation} \label{eq:adj}
0 \to T\cM \to T\cA \to \cE_1 \to 0 \; .
\end{equation}
We will use this to calculate cohomologies and Chern classes of $X$.
\item The Koszul sequence, which for a codimension one submanifold $\cM$ reads 
\begin{align}
\label{seq:koszul}
    0  \rightarrow \cE_1^* \rightarrow \cO_\cA \rightarrow \cO_A|_\cM \rightarrow 0\; .
\end{align}
Here $\cE_1^*$ is the dual of the normal bundle. We will use this sequence, and tensor products of it, to compute cohomologies of $X$ and line bundles over $X$.
\end{itemize}

\noindent{In appendix \ref{ap:topology}, we apply these tools to compute the topological quantities of relevance for the gCICY manifolds $X_1, X_2$. In particular, this confirms that 
\[\chi(X_1) =-80, \, h^{1,1}(X_1)=5, \, h^{2,1}(X_1)=45 \text{ and } \chi(X_2) =-48, \, h^{1,1}(X_2)=5, \, h^{2,1}(X_2)=29 \,\]
as stated in the configuration matrices \eqref{eq:gCICYs}.}

\subsubsection*{Line Bundle Cohomology}

We compute the dimension of line bundle cohomologies $h^\bullet(X, L)$ using a stepwise implementation of the computational tools defined in the last subsection. Tensoring the Koszul resolution with $L$ yields
\begin{align}
    0 \rightarrow \cE_1'^* \otimes L|_\cM \rightarrow  L|_\cM \rightarrow L|_X \rightarrow 0.
\end{align}
The relevant entries in the associated long exact cohomology sequence are again computed via the Koszul resolution in \eqref{seq:koszul} over the ambient space $\cA$ with
\begin{align}
    &0 \rightarrow \cE_1^* \otimes L|_\cA \rightarrow  L|_\cA \rightarrow L|_\cM \rightarrow 0 \\
    &0 \rightarrow \cE_1'^* \otimes \cE_1^* \otimes L|_\cA \rightarrow \cE_1'^* \otimes L|_\cA \rightarrow \cE_1'^* \otimes L|_\cM \rightarrow 0.
\end{align}
The cohomologies over $\cM$ can be computed using the Bott-Borel-Weil theorem and the well established methods described in \cite{Anderson:2008ex,Hubsch:1992nu}. When computing maps between non trivial cohomology entries on $\cM$ one has to be extra careful using the construction of the rational sections $\cE_1'$ given in equation \eqref{eq:rsection}. Furthermore, one cannot simply use the tensor representations arising from the Bott-Borel-Weil theorem, when there are already maps involved in the cohomology computations on $\cA$. It is then required to project the tensors onto the respective kernel or image using the defining monomials of $\cE_1$. On regular CICYs we matched this stepwise computation of cohomologies to the results obtained by applying the Leray spectral sequences arising from a long exact Koszul resolution. We note that there exists a refined version of the Koszul resolution \cite{Berglund:2016yqo} which possibly allows to side step our stepwise computation. However, this was not used in the line bundle computations in this paper.

\subsection{Kähler Cone}
\label{sec:kaehler}

The Kähler cone consist of $J \in H^2(X)$ satisfying 
\begin{align}
    \int_X J \wedge J \wedge J > 0, \quad \int_S J \wedge J > 0, \quad \int_C J > 0   \label{eq:posconstr}
\end{align}
for all $S,C \subset X$ homologically non-trivial, reduced proper surfaces and curves in $X$. Here each $J$ is Poincaré dual to an ample divisor, and thus the Kähler cone is the cone of ample divisors. Moreover, this constraint  shows that the curves $C$ lie in the Mori cone (the cone of effective curves), which is the dual of the Kähler cone.

The K{\"a}hler cones for the two gCICYs were studied following a four-step prescription described in Ref. \cite{Anderson:2015iia}. First, the Mori cone  of the hypersurface $\cM$ defined by the polynomial section is found by studying the cohomology of the line bundles on the ambient space and the Koszul sequence. 
Second, the generators of the Mori cone are used to construct homologically non-trivial, reduced submanifolds and equations \eqref{eq:posconstr} are checked directly to find the K{\"a}hler cone for $\cM$.
In steps three and four, steps one and two are repeated including the rational section to obtain the K{\"a}hler cone for the gCICY.
More generally, this algorithm could be extended for the calculation of the K{\"a}hler cone for gCICYs with any number of polynomial or rational constraints by repeating steps one and two for every constraint.

We present the details of this analysis, for the gCICY manifold $X_2$, in appendix \ref{apx:KahlerAppendix}. What we find, for both gCICY geometries, is that the Kähler cone is contained in the positive orthant $t^i > 0$ of the Kähler moduli space.

\subsection{Freely Acting Discrete Symmetries}
\label{sec:freely}

The freely acting discrete symmetries of ordinary CICY manifolds have been classified in Ref.~\cite{Braun:2010vc}. This classification  is based on first charting the symmetries of the ambient space, and then construct CICY-defining polynomials that respect these symmetries. A similar classification should, as already noted in \cite{Anderson:2015iia}, be possible for gCICY manifolds. In this section we will take a first step towards this goal, by showing that the two gCICYs $X_1$ and $X_2$ admit freely acting $\mathbb{Z}_2$ symmetries.   Our analysis closely follows that of Ref.~\cite{Candelas:2008wb}, with modifications needed to tackle the added technical challenge that the  rational constraints pose to establishing discrete symmetries (freely acting or not). 

The process of finding a freely acting symmetry on a gCICY can be boiled down to the following three steps: 
\begin{enumerate}
    \item Guess an action $g$, generating a (not necessarily freely acting) discrete  symmetry $\Gamma$ on the ambient space coordinates and gCICY constraints.
    \item Find invariant defining hypersurface equations and check for smoothness using the techniques described in section \ref{sec:smoothness}.
    \item Check that  the Calabi Yau misses all the fixed points of $\Gamma$. 
\end{enumerate}

A good guess for a freely acting $\bZ_2$ symmetry is the following action on the ambient space coordinates
\begin{align}
  g: y_{i,j} \rightarrow (-1)^{j+1}y_{i,j} 
  \label{eq:grpact}
\end{align}
for $i=0,\dots,4,\ j=0,1$ and on the constraints with 
\begin{align}
    g_{\text{constr}}: p_i \rightarrow (-1)^i p_i
	\label{eq:grpactpoly}
\end{align}
for $i=0,1$ and $p_1$ corresponding to the rational section.

In the following subsections we will show that this action on the ambient space describes a discrete symmetry on the two gCICYs presented in this paper.
After constructing the symmetric polynomial constraint we demonstrate how the rational constraint can be made symmetric with respect to $g$. We then show how this symmetry is maintained for the rational section over all of its continuations where the denominators vanish. Finally, we confirm that these constraints remain smooth manifolds and miss all fixed points.
\subsubsection*{X1}
To establish the presence of a freely acting discrete symmetry the gCICY is to be constructed using polynomial and rational constraints symmetric under the action of $g$ and $g_{\text{constr}}$.
Because there are no significant constraints on the polynomial section, this can be chosen to be the most general symmetric polynomial with respect to $g$ with an overall factor given by $g_{\text{constr}}$.
Specifically, the polynomial constraint \eqref{eq:polyconstr1} is made symmetric by choosing $d_i$ such that $g(d_i)=(-1)^{i+1}d_i$.

A general rational constraint can then be obtained by using the polynomial constraint \eqref{eq:polyconstr1} according to the standard procedure described in section \ref{sec:twogcicys}.
The rational constraint is made symmetric by further constraining the free coefficients $\alpha_2$ and $\alpha_3$ in equations \eqref{eq:ratsec1R0} and \eqref{eq:ratsec1Ri} to be equal. It is noteworthy that under the transformation $g$ the denominators $\Delta_2$ and $\Delta_3$ are exchanged (up to a minus sign). Therefore, the regions $R_2$ and $R_3$ are also exchanged. The symmetry for the rational constraint however is preserved as $m_{2}\to -m_{3},\ m_{3}\to -m_{2}$ and $N_{2}\to N_{3},\ N_{3}\to N_{2}$.

\subsubsection*{X2}
Similarly we find for the second gCICY that the polynomial constraint is the same as \eqref{eq:polyconstr1} with the exception that $d_{i}$ do not depend on the fourth $\bP^{1}$.
The rational constraint takes the general form:
\begin{equation}
  q_{0}=\sum_{i=1}^{3}\frac{N_{i}}{\Delta_{i}}\left( \alpha_{i}y_{3,0}^{2}+\beta_{i}y_{3,0}y_{3,1}+\gamma_{i}y_{3,1}^{2} \right)
\end{equation}
where $\alpha_{i},\ \beta_{i}$ and $\gamma_{i}$ are constants and $N_{i}$ and $\Delta_{i}$ are as before. To make this constraint symmetric under the action of $g$, one must impose: $\beta_{i}=0,\ i=1,2,3,\ \alpha_{2}=\alpha_{3}$ and $\gamma_{2}=\gamma_{3}$. The numerators $m_{1},\ m_{2}$ and $m_{3}$ also take the same form, therefore, with the restrictions above the rational constraint is symmetric on all four regions $R_{i}$.

\subsubsection*{Smoothness and fixed points}

The restrictions just imposed on the complex moduli are needed in order to make the gCICY manifolds symmetric with respect to the group action on the ambient space. However, they may compromise the smoothness of the geometry. We must therefor perform a secondary smoothness check for the two geometries. In both cases we find that the Gr{\"o}bner basis techniques described in section \ref{sec:smoothness} give positive results for the smoothness of the gCICYs.

We then proceed to check that any fixed points of the discrete symmetry are avoided on the gCICY hypersurface, as required to obtain a smooth quotient manifold $X_i/\Gamma$. This was analyzed the same way as smoothness.
For each $\mathbb{P}^1$ factor of the ambient space, the coordinates of the fixed points of the symmetry \eqref{eq:grpact} are, up to projective equivalence, $\left( y_{i,0},y_{i,1} \right)=\left( 0,1 \right)$ and $\left( 1,0 \right)$. Hence, there are 32 fixed points. A Gr{\"o}bner basis calculation was performed for each fixed point to check that it does not lay on the gCICY. These calculations show that all fixed points for both gCICYs are removed from the surface, and hence the gCICYs are fixed point free.  We conclude that the quotient manifolds resulting from dividing the gCICY by the $\mathbb{Z}_2$ symmetry are smooth. 

\section{Scan Results}
\label{sec:results}

In this section we present our results of systematic scans over the two geometries $X_1$ and $X_2$. First, we go briefly through the computations of one particular standard-like model on the second geometry. In the second part we compare the number of models found to similar CICY geometries. To reduce the technical complexity of our discussions, we focus exclusively on the properties of the “upstairs” models, where the discrete symmetry has not been quotiented out (cf. [19]).

\subsection{One Example Model}
\label{sec:example}

In this section we illustrate the constraints presented in section \ref{sec:lbmodels} with the help of one explicit example. 
Consider the manifold $X_2$ and the following sum of line bundles
\begin{align}
    V &= \cO_{X_2} (-1,0,1,0,-1) \oplus \cO_{X_2} (-1,1,0,0,-2) \oplus \cO_{X_2} (0,0,1,-2,2) \nonumber\\
    & \qquad \oplus \cO_{X_2} (1,-1,-1,1,2) \oplus \cO_{X_2} (1,0,-1,1,-1).
\end{align}
V has a vanishing first Chern class, and its second Chern class is given by
\begin{align}
    c_2(V) = (18,20,16,8,14) \quad < \quad c_2({X_2}) = (24,24,24,24,24) 
\end{align}
which is strictly larger than 0 thus satisfying Bogomolov bound and Bianchi identity \eqref{eq:bianchi}. Using equation \ref{eq:slope} we can show numerically that $V$ has vanishing slope somewhere in the positive orthant of the K\"ahler moduli space as required for stability. Furthermore, the rank of $V$ is 4, such that all $U(1)-$bosons acquire a mass through the Green-Schwarz mechanism.
% particle content
The index of $V$ is
\begin{align}
    \chi ( V) = -6 = -3 \cdot |\mathbb{Z}_2|
\end{align}
leading to a three generation model. Next we check the cohomology computations explicitly to verify that there are no anti-families and at least one Higgs doublet.
In order to find the number of anti generations we proceed by applying the Koszul resolution \eqref{seq:koszul} step-wise and utilizing the Bott-Borel-Weil theorem to compute coholomogies of line bundle on products of projective spaces. After some short exact sequence chasing we arrive at the following result:
The only non vanishing Hodge number of $V$ arises from the line bundle $L_3 = \cO_{X_2} (0,0,1,-2,2)$ with
\begin{align}
    H^1(X_2, L_3) \cong H^1(\cM_2, L_3) \cong H^1(\cA, L_3 \otimes \cE_2^*)
\end{align}
which has dimension $h^1(X_2, L_3) = 6$.
Similarly we find the following non vanishing cohomologies for $V \wedge V$% and $V^* \wedge V^*$:
\begin{align}
    H^1(X_2, L_1 \otimes L_2) & \cong H^1(\cM_2, L_1 \otimes L_2 ) \nonumber \\
    & \cong \text{Im} \left( H^1(\cA, L_1 \otimes L_2 \otimes \cE_2^*) \rightarrow H^1(\cA, L_1 \otimes L_2 ) \right) \nonumber \\
    & \Rightarrow h^1(X_2, \cO_{X_2} (-2,  1,  1,  0, -3)) = 2 \\
    H^2(X_2, L_1 \otimes L_2) & \cong H^3(\cM_2, L_1 \otimes L_2 \otimes \cE_2'^*) \cong H^3(\cA,L_1 \otimes L_2 \otimes \cE_2'^*) \nonumber \\
    & \Rightarrow h^2(X_2, \cO_{X_2} (-2,  1,  1,  0, -3)) = 2\\
    H^1(X_2, L_1 \otimes L_4) & \cong H^1(\cM_2, L_1 \otimes L_4) \cong H^1(\cA, L_1 \otimes L_4) \nonumber \\
    & \Rightarrow h^1(X_2, \cO_{X_2} (0,  0,  0,  1, -2)) = 2 \\
    H^1(X_2, L_2 \otimes L_3) & \cong \text{Ker} \left( H^2(\cM_2, L_2 \otimes L_3 \otimes \cE_2'^*) \rightarrow H^2(\cM_2, L_2 \otimes L_3 ) \right) \nonumber \\
    & \cong \text{Ker} \left( H^2(\cA, L_2 \otimes L_3 \otimes \cE_2'^*) \rightarrow H^3(\cA, L_2 \otimes L_3 \otimes \cE^*_2 ) \right)\nonumber \\
    & \Rightarrow h^1(X_2, \cO_{X_2} (-1,  1,  1, -2,  0)) = 4
\end{align}
leading to a single Higgs doublet surviving in the downstairs spectrum and three massless fermion generations. A simple index computation also shows that the triplet constraint \eqref{eq:higgs} can be verified for all pairs $L_a \otimes L_b$. Hence there is no complete {\bf 5} surviving in the downstairs theory and the Higgs triplets can be projected out. Finally, we note that there are 9 families of singlets arising from $V \otimes V^*$, yielding
\begin{align}
    4_{2,5} + 8_{3,1} + 20_{3,2} + 8_{3,4} + 16_{3,5} + 8_{4,1} + 16_{4,2} + 6_{5,1} + 4_{5,4} = 90_{V \otimes V^*}
\end{align}
in total 90 bundle moduli in the upstairs spectrum. Here, the subscripts denote the product of line bundles and singlet families.

\subsection{All Standard-Like Models}
\label{sec:statistics}

\begin{algorithm}[t]
\SetAlgoLined
\KwResult{Creating all line bundle models on X for a given $q_{max}$ in $L_4$.}
 initialize $L_0 = []$\;
 \For{$L \in $ all line bundles given $q_{max}$}{
 \uIf{satisfies slope condition (\ref{eq:slope})}{
 \uIf{$h^\bullet (X, L) = (0,x,0,0)$ with $x \in \{0, |\Gamma|, 2 \cdot |\Gamma|, 3 \cdot |\Gamma|\}$}{
 add $L$ to $L_0$\;
 }
 }
 }
 \For{$i=1; i < 5; i++$}{
  initialize list of tuples $L_i = []$\;
  \While{$\exists t_l \in $ all combinations ($t_{i-1;j}, l_{0;k}$)}{
  \uIf{$t_l$ satisfies conditions from section (\ref{sec:lbmodels})}{
    add $t_l$ to $L_i$\;
  }
  }
  remove duplicates in $L_i$\;
 }
 \caption{Simplified sketch of the scanning algorithm for a given $q_{max}$.}
 \label{alg:algo}
\end{algorithm}

\begin{figure}
\centering
\includegraphics[scale=0.65]{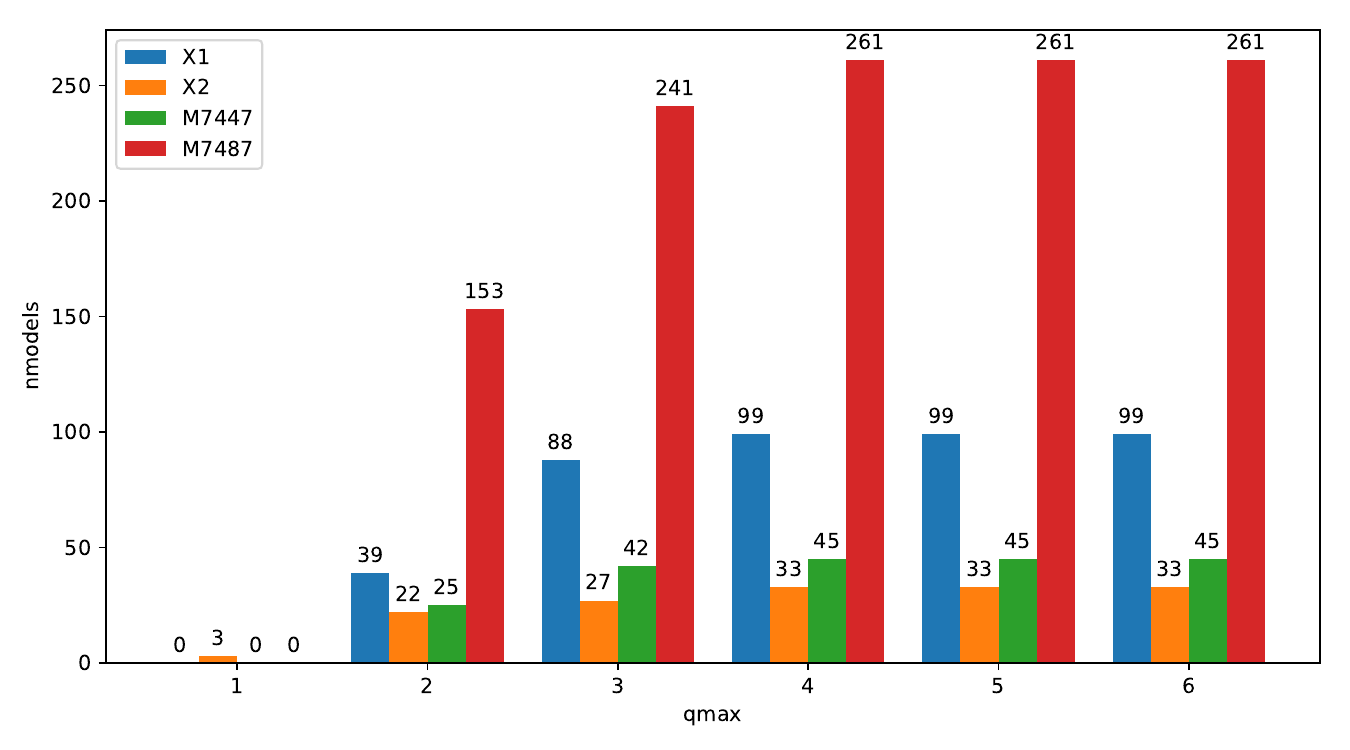}
\caption{Number of models found plotted against $q_{max}$ for the two geometries studied in this paper and the CICYs with numbers 7447 and 7487.}
\label{fig:plot}
\end{figure}

We scan systematically for all line bundle models on the two manifolds $X_1$ and $X_2$. The sketched algorithm \ref{alg:algo} employed is the same as outlined in section 5 of \cite{Anderson:2013xka} and can be found online \cite{gCICYtoolkit}. 
We repeat the algorithm with increasing values for $q_{max}$ until we can no longer find new models for two further increments. {We end the algorithm by removing duplicate line bundles, related by a permutation of charges on projective spaces that have the same constraints\footnote{ E.g. for $X_1$ this removes line bundle models related by permutations of the first four charges, for $X_2$ the first three.}. } {The results of this scan are shown in Figure \ref{fig:plot}}. In the end we find 99 models on the first and 33 models on the second gCICY, leading to a total of 132 standard-like models. The explicit line bundle sums are found in Table \ref{tab:models1} and Table \ref{tab:models2}. We will now put this result into perspective, and comment on uniqueness of the found models, by comparing our result to previous heterotic SLM constructions.

On both gCICYs we find that the maximal charge $q_{max} = 4$ that occurs is in line with the two CICY geometries. It would be illuminating to derive this bound on the maximal charge as was done for the tetraquadric in \cite{Buchbinder:2013dna}. We also see that the highest increase in models found occurs at $q_{max} = 2$, which is consistent with the results of \cite{Anderson:2013xka}. There is one more observation to be made: $X_2$ admits three models with $q_{max} = 1$. This is a property that for CICYs is only observed at $h^{1,1}(X) \geq 6$, again emphasizing qualitative differences between the two classes of CY manifolds.
Moreover, we can rest assured that the 33 models found on $X_2$ are new heterotic line bundle SLMs. Indeed, the Hodge pair (5,29) of $X_2$ does not appear in the CICY list, and while there are four polytopes in the Kreuzer-Skarke list with these Hodge numbers, they have not been used in the construction for SLMs \cite{Braun:2017feb,He:2013ofa}.

\subsubsection*{Comparison to data on CICYs}  

It is interesting to compare our results to the ones found on regular CICYs. The following two geometries
\begin{align}
\label{eq:CICYs}
      M_{7447}\in\left[ \begin{array}{c||cc}
		1 & 1 & 1 \\
		1 & 1 & 1 \\
		1 & 1 & 1 \\
		1 & 1 & 1 \\
		1 & 1 & 1 
	\end{array}\right]_{-80}^{5,45} \qquad \text{and} \qquad 
    M_{7487}\in\left[ \begin{array}{c||cc}
		1 & 0 & 2 \\
		1 & 1 & 1 \\
		1 & 1 & 1 \\
		1 & 1 & 1 \\
		1 & 1 & 1 
	\end{array}\right]_{-80}^{5,45}
\end{align}
are, according to Wall's theorem \cite{wall1966classification}, homotopically equivalent to $X_1$\footnote{Two Calabi-Yau three folds are said to be homotopically equivalent, if all their topological invariants such as $c(X), d_{ijk}, h^\bullet$ match.}. In \cite{Anderson:2013xka} it has been shown that these CICYs admit respectively 93 and 459 heterotic line bundle models. Excluding models with anti-generations and not satisfying the Higgs constraints \eqref{eq:higgs} they still admit 45 and 261 heterotic SLM. This leads to the interesting observation that, while these two manifolds are homotopically equivalent, they admit different number of SLM realisations, showing that the number of SLMs depends strongly on the embedding of the CY. This difference can largely be explained by the broken permutation symmetry in the first row of the configuration matrix of $M_{7487}$. This leads to less redundant models in this case, compared with of $M_{7447}$. Since the configuration matrix of $X_1$ in \eqref{eq:gCICYs} also has the last row breaking a permutation symmetry of the five projective spaces, we would expect the number of models found to be closer to $M_{7487}$. 

\begin{figure}
\centering
\begin{minipage}{0.49\textwidth}
\includegraphics[width=\textwidth]{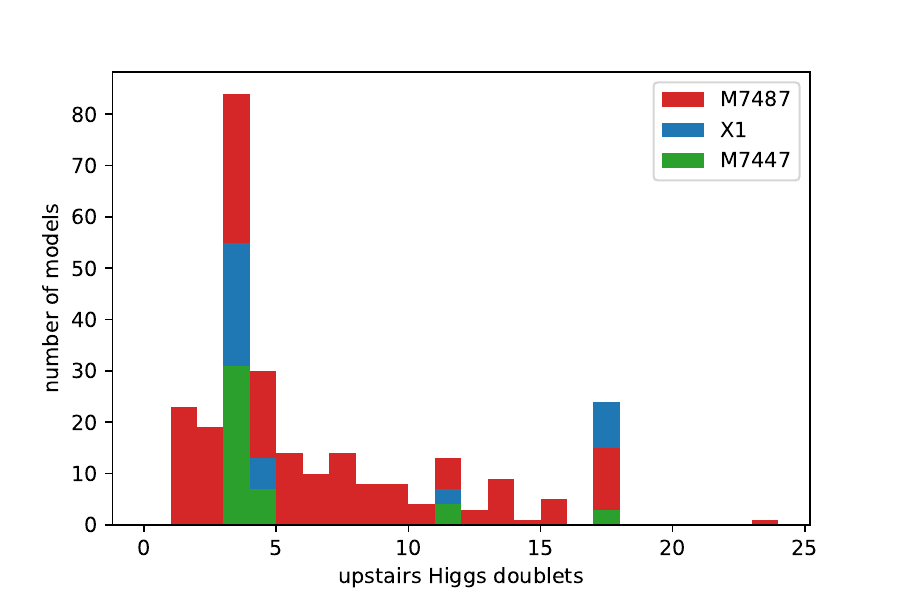}
\end{minipage}
\begin{minipage}{0.49\textwidth}
\includegraphics[width=\textwidth]{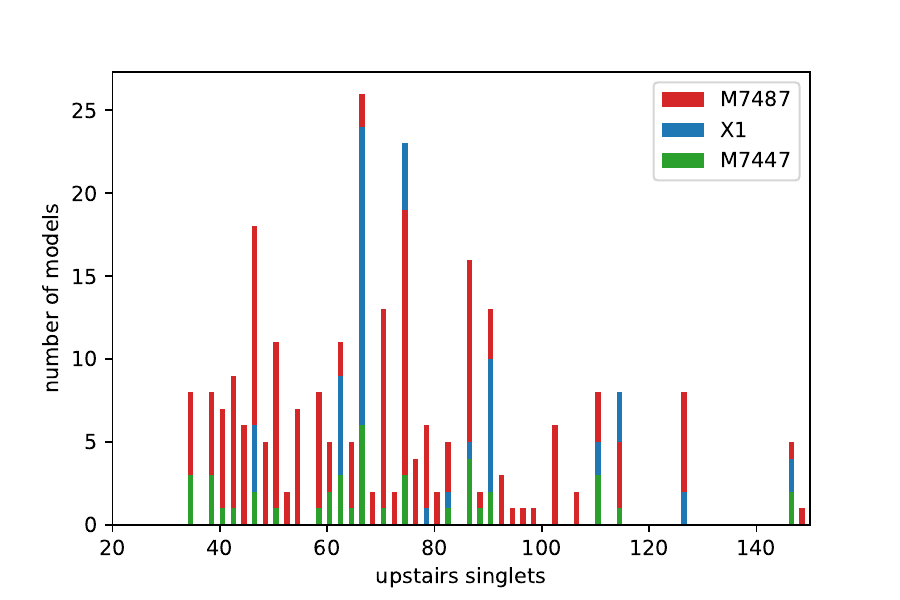}
\end{minipage}
\caption{On the left are the number of Higgs doublets in the upstairs theory for each model on the three equivalent CY geometries. On the right are the number of bundle singlets.}
\label{fig:higgs}
\end{figure}

Figure \ref{fig:plot} shows all models found on the four geometries for increasing $q_{max}$. The 99 models on $X_1$ are about twice as much as the 45 models on $M_{7447}$ but do not reach the 261 models found on $M_{7487}$. This already indicates that the gCICY presentation is comparably efficient in providing heterotic SLMs.
It is worth taking a closer look at the models found on these three equivalent manifolds. This will tell us whether we have found truly new SLMs in the gCICY description. We thus plot, in Figure \ref{fig:higgs}, the number of Higgs doublets and vector bundle singlets in the upstairs spectrum. The two plots suggest that the models found on $X_1$ and $M_{7447}$ are, to a large degree, a subset of the models found on $M_{7487}$. 

There are, however, a few outliers. From the Higgs doublet plot on the left we see that there are more models with 17 Higgs doublets on $X_1$ (24) than on the two CICY  realisations (15 and 3, respectively). Comparing the set of topolgical invariants, given by $h^2(V), h^{1,1}(V \otimes V^*)$ and $c_\bullet(V)$, we find that all combinations match to existing models on $M_{7487}$. This also holds true for models found on $M_{7447}$.
Similar observations can be made for the singlet peaks at $h^{1,1}(V \otimes V^*) = \{74, 114\}$ of $X_1$ models. They come with respectively 3 and 17 Higgs doublets, but again the combination of Higgs doublets, singlets and $c_\bullet(V)$ is not unique to the $X_1$ description. In conclusion it appears that from the upstairs perspective all the models found on $X_1$ (and also on $M_{7447}$) are merely different realisations of models already found on $M_{7484}$.

\section{Discussion and Outlook}
\label{sec:outlook}

In this paper, we have considered two gCICY manifolds, that were first constructed in Ref.~\cite{Anderson:2015iia}, as internal manifolds for heterotic line bundle standard-like models. We proved that both geometries admit freely acting $\mathbb{Z}_2$ symmetries, and thus may be used to construct smooth quotient manifolds with a non trivial fundamental group. A systematic scan of line bundle sums over these manifolds revealed that the they respectively admit 33 and 99 heterotic SLM, cf. figure \ref{fig:plot}. These putative standard models satisfy several low energy constraints, such as exactly three fermion generations, a unifying GUT group, no Higgs triplets, and at least one Higgs doublet.\footnote{Our analysis has been limited to the ``upstairs" model, in the parlance of \cite{Anderson:2013xka}, and we have left the explicit construction of equivariant line bundle cohomologies needed for the ``downstairs" model that prevail on the quotient manifold for future work.}  {While these are non-trivial tests, they are but the first step towards the construction of  realistic string models for particle physics, and below we discuss  directions for future studies that are needed to fully establish the relevance of these models. Furthermore, having found a number of models for a gCICY in the same order of magnitude as that for an equivalent CICY presentation, we can establish a reasonable effectiveness of gCICY presentations as compared to regular CICYs.} 

A natural extension of our work would be to extend our systematic scan to more gCICY examples. This, however, requires a more detailed classification of favourable and smooth gCICY geometries, where determining the K\"ahler cone is a particularly important non-trivial step.  After such a classification one should proceed, using the methods described in this paper, by finding all freely acting symmetries on this set of manifolds as was done for the CICYs \cite{Braun:2010vc}, and ideally also determine the first fundamental group of the gCICY manifolds before quotienting, so that a complete analysis of Wilson line symmetry breaking can be performed. Then, finally one may perform systematic scans for realistic vacua, as done for the two example manifolds in this paper. Here, it would be very useful to develop some systematic/algorithmic methods to determine the rational sections, such that the manifolds do not have to be treated on a case by case basis.

The methods we have developed in this paper will likely be of relevance outside the realm of heterotic line bundle model building. Line bundle cohomologies are a stepping stone for the computation of cohomologies of monad and extension bundles, which could be interesting to explore also on gCICY geometries. Recently, it has also been observed that cohomologies of line bundles over CY threefolds may be captured by certain analytic cohomology formulae \cite{Buchbinder:2013dna,Constantin:2018hvl,Klaewer:2018sfl,Larfors:2019sie,Brodie:2019dfx,Brodie:2020wkd,Brodie:2020fiq}.\footnote{Related discussions, and mathematical proofs of these formulae, exist for twofolds \cite{Brodie:2019ozt,Brodie:2019pnz}.} When explicitly constructing sections needed for the analysis of K\"ahler cones, we have seen tantalizing indications of such structures in the two gCICY threefolds studied here.

Returning to the setting of heterotic line bundle building,  there is scope to analyse mass terms and couplings in the SLM we have found, in order to establish their phenomenological quality. Of particular interest are the Yukawa couplings, which may be partly determined using algebraic geometry. Here,  the gCICY geometry offers an interesting test of recent vanishing theorems for Yukawa couplings in the resulting models. In Ref.~\cite{Blesneag:2015pvz,Blesneag:2016yag} a particular pattern was noticed for Yukawa couplings in heterotic line bundle models on CICYs. In particular, the authors of \cite{Blesneag:2015pvz,Blesneag:2016yag} assign an integer ``type" for each matter field in the model, corresponding to the ambient space cohomology that the field originates from. The theorem then states that the sum of types must be larger than the ambient space dimension for there to be non trivial couplings between fields. This is interesting for gCICYs since the negative numbers in the configuration matrix \eqref{eq:gCICYs} generally lead to higher type (i.e. the Bott-Borel-Weil theorem would relate the matter field to higher ambient space cohomologies), so one would expect more Yukawa couplings to be allowed on gCICYs than on CICYs.\footnote{It would also be interesting to explore other patterns for Yukawa couplings in the gCICY setting. For example, the fibration structure of CY manifolds lead to powerful selection rules \cite{Braun:2006me}, which have recently been studied systematically for CICY manifolds \cite{Gray:2019tzn}. We expect similar results in the gCICY case.} 

{In a similar manner, it would be interesting to explore moduli stabilising mechanisms and non-perturbative corrections  in heterotic compactifications in the gCICY regime. In the perturbative regime, moduli stabilisation may arise through interaction between bundle and geometric degrees of freedom, as recently demonstrated in Refs~\cite{Anderson:2009nt,Anderson:2011ty,Anderson:2014xha, delaOssa:2014cia,Garcia-Fernandez:2015hja,delaOssa:2016ivz,delaOssa:2017pqy}.  In brief, the mechanism behind this stabilisation is related to certain jumps in dimension of line bundle cohomology classes.  Thus, just as for the Yukawa coupling, one might expect that the rational defining relations of gCICYs may affect the ability to stabilise moduli. Furthermore, non-perturbative corrections are of essence for the stability of heterotic models. Recently, it has been shown that the assumptions underlying the vanishing theorems of Ref.~ \cite{Distler:1986wm,Distler:1987ee,Silverstein:1995re,Beasley:2003fx} do not necessarily apply in generic heterotic compactifications \cite{Buchbinder:2018hns,Buchbinder:2019hyb}. In the latter references, examples of non-vanishing  non-perturbative superpotentials were found among both CICY and elliptically fibered CY manifolds. It would be rewarding to extend this analysis to the gCICY setting, which indeed manifestly violate the assumptions of \cite{Beasley:2003fx}. However, since fully exploring these effects go beyond the restricted  line bundle setting, they are left as intriguing topics for future studies.}

Finally, we want to state that while our work shows that heterotic model building on gCICYs is possible, it also shows that it comes with some increased difficulties compared to regular CICYs. Hence, it would be rewarding to explore alternative paths to analyse gCICY compactifications. Recently, promising results have been obtained from employing techniques from Data Science to solve geometrical problems. Related to heterotic line bundle models, success was reported in finding realistic configurations on CICYs using reinforcement learning \cite{Larfors:2020ugo} and identifying clusters of models using auto-encoders \cite{Deen:2020dlf,Otsuka:2020nsk}. These methods may also perform well in the gCICY setting. {Machine learning has also recently provided novel approximations of CY metrics \cite{Ashmore:2019wzb,Douglas:2020hpv,Anderson:2020hux}, that may provide new avenues to explore gCICY geometry.} Furthermore, it would be of great interest to apply machine learning to identify discrete symmetries, and analyse the smoothness of manifolds. {We hope to return to these topics in the future.}

%%%%%%%%%%%%%%%%
\section*{Acknowledgements}

The authors would like to thank P.~Candelas and A.~Constantin for clarification regarding freely acting symmetries, and L.~Anderson and J.~Gray for discussions pertaining to line bundle cohomology computations. This work is financed by the Swedish Research Council (VR) under grant numbers 2016-03873 and 2016-03503. D.P. is supported by the NWO vidi grant (number 016.Vidi.189.182).

%%%%%%%%%%%%%%%%%%%%%%%
%\newpage
\begin{appendix}

\section{Computation of Topological Invariants}

\label{ap:topology}
In this appendix, we apply the computational tools introduced in section \ref{sec:sequences} to  compute topological quantities of the gCICY geometries $X_1$ and $X_2$.

{\bf Triple intersection numbers}: The intersection numbers of divisors on $X$ may be computed as
\begin{align}
    d_{ijk} = \int_X J_i \wedge J_j \wedge J_k  = \int_\cA J_i \wedge J_j \wedge J_k \wedge \mu_X \wedge \mu_\cM \; ,
\end{align}
where we introduce the measures $\mu_\cM$ and $\mu_X$ as in Ref.~\cite{Anderson:2015iia}, which are, respectively, Poincaré duals of $\cM$ and $X$.

For $X_{1}$ the nonzero triple intersection numbers, up to symmetric permutation of the indices, are:
\begin{equation}
  d_{123}=d_{124}=d_{125}=d_{134}=d_{135}=d_{145}=d_{234}=d_{235}=d_{245}=d_{345}=2.
\end{equation}
For $X_{2}$ we find:
\begin{equation}
  \frac{1}{3}d_{123}=d_{124}=d_{125}=d_{134}=d_{135}=d_{145}=d_{234}=d_{235}=d_{245}=d_{345}=2.
\end{equation}

{\bf Chern classes}: 
By applying the adjunction formula repeatedly for the succession of manifolds ${\cal M} \supset {\cal M}_1 \supset {\cal M}_2...\supset {\cal M}_{L-1} \supset X$, we may compute the Chern classes of $X$ in a stepwise manner. In particular, we may use this procedure to  compute \cite{Anderson:2015iia}
\begin{equation}
c_1^i(X) = (n_i +1)-\sum_{j=1}^K a_j^i - \sum_{j=1}^L b_j^i \; ,
\end{equation}
where $c_1 = c_1^i J_i$, and $J_i$ is a basis for the K\"ahler forms on $X$.
Since the vanishing of the first Chern class is required for $X$ to be a Calabi--Yau manifold, this reproduces the familiar CY condition that the sum of the entries in each line in the configuration matrix must equal the dimension of the projective space plus one. For the manifolds $X_{1,2}$ we further find that:
\be
c_{2}(X_{1,2}) = \left(24,24,24,24,24\right), 
\ee
where $c_{2}$ is the second Chern class in vector form.

{\bf Euler characteristic}: Since $X$ is a complex manifold, its Euler characteristic ${\chi}$  is given by the Chern--Gauss--Bonnet theorem as the integral of the top Chern class
\begin{equation}
\chi(X) = \int_X c_3(X) =  \int_\cA J_i \wedge J_j \wedge J_k \wedge \mu_X \wedge \mu_\cM \; , 
\end{equation}
where in the lift to an ambient space integral we specialise to codimension (1,1) gCICYs, and again use the measures $\mu_\cM$ and  $\mu_X$ to pullback the integral to $\cA$. This ambient space integral can be expanded in terms of the entries of the configuration matrix, see \cite{Passaro:2020}, but we will not need the expression here. We readily find that $\chi(X_1) =-80$ and $\chi(X_2) =-48$, as stated in the configuration matrices \eqref{eq:gCICYs}.

{\bf Hodge numbers}:
Finally, we will need to compute the non-trivial Hodge numbers  $h^{(1,1)},h^{(2,1)}$ of $X_{1,2}$ , as well as Hodge numbers of line bundles over them. These computations rely on repeatedly applying the tools listed in section \ref{sec:sequences}. Here, we illustrate these computations by determining the Hodge numbers for $\cM_1$, referring to  \cite{Passaro:2020} for more details.  

In this case $\mathcal{A}=\left( \mathbb{P}^{1} \right)^{\oplus 5}$ is the ambient space, $\mathcal{E}_1=\mathcal{O}\left( 1,1,1,1,3 \right)$ is the normal bundle to $\mathcal{M}_{1}$ in $\cA$,  and $\mathcal{E}_1'=\mathcal{O}_{\mathcal{A}}\left( 1,1,1,1,-1 \right)$ is the normal bundle of $\mathcal{X}_{1}$ in $\mathcal{M}_{1}$. $\cE_1^*$ denotes the conjugate of the normal bundle.
\begin{enumerate}
\item Use the Koszul sequence \eqref{seq:koszul}, with respect to $\mathcal{M}_{1}\subset \mathcal{A}$ tensored with $\mathcal{E}_1$ to compute $h^{\bullet}\left( \mathcal{M}_{1},\mathcal{E}_1 \right)$ in terms of $h^{\bullet}\left( \cA,\mathcal{E}_1 \right)$ and $h^{\bullet}\left( \cA,\cO \right)$, which may in turn be computed using the Bott-Borel-Weil theorem. This leads to $$h^{\bullet}\left( \mathcal{M}_{1},\mathcal{E}_1 \right)=\left( 63,0,0,0,0 \right).$$
\item  Use the Koszul sequence tensored with $T\cA$ to compute in terms of $h^{\bullet}\left( \cA,T\cA \right)$ and $h^{\bullet}\left( \cA,\mathcal{E}_1^* \otimes T\cA \right)$. These ambient space cohomologies may be computed using the Euler sequence \eqref{eq:euler} and the Bott-Borel-Weil theorem. 
This gives $$h^{\bullet}\left( \cM_1,T\mathcal{A} \right)=\left( 15,0,0,0,0,0 \right).$$
\item Use the adjunction formula \eqref{eq:adj} and the associated long exact sequence in cohomology to determine the cohomology group dimensions of $T\cM_1$ in terms of of $h^{\bullet}\left( \mathcal{M}_{1},\mathcal{E}_1 \right)$ and $h^{\bullet}\left( \cM_1,T\cA \right)$. We find $$h^{\bullet}\left( \mathcal{M}_{1},T\mathcal{M}_{1} \right)=\left( b,48+b,0,0,0 \right).$$ where the undetermined parameter $b$ reflects the non-shortness of the cohomology sequence.
\end{enumerate}
The steps are then repeated to determine the $TX_1$-valued cohomology of $X_1$. Glossing over some technical details, this allows to determine the Hodge number 
\be h^{(1,1)}(X_1)= h^2(X_1, T X_1)=5 \; . \ee 
which shows that the manifold is K\"ahler favourable, i.e. that $h^{(1,1)}(X_1) = h^{(1,1)}(A)$. Then, using the Euler characteristic, we may conclude that 
\begin{equation}
  h^{2,1}\left( X_{1} \right)=-\frac{1}{2}\chi+h^{1,1}\left(X_{1} \right)=45 \; 
\end{equation}
in agreement with the configuration matrix \eqref{eq:gCICYs}.

The same computation for gCICY $X_2$ reproduces 
\begin{gather*}
  h^{1,1}=5 \quad h^{2,1}=29.
\end{gather*}

\section{Computation of K\"ahler Cone}

\label{apx:KahlerAppendix}

In this appendix, we discuss the Kähler cone of $X_2$. Similar results pertain to $X_1$. Recall that the gCICY $X_{2}$, the intermediate hypersurface $\cM$ and the ambient space $\cA$ are given by
$$X_2=\left[\begin{array}{c||c|c} 1 & 1 & 1\\ 1 & 1 & 1 \\ 1 & 1 & 1 \\ 1 & 0 & 2 \\ 1 & 3 & -1\end{array}\right],\ \cM=\left[\begin{array}{c|c} 1 & 1\\ 1 & 1 \\ 1 & 1 \\ 1 & 0 \\ 1 & 3 \end{array}\right],\  \cA=\left(\bP^1\right)^{\otimes 5}.$$ 

The class of effective divisors of $\cA$ is defined as the class $D=\sum_{i=1}^{5}a_{i}H_{i}$, where $a_{i}\ge0$ and $H_{i}$ are the hyperplanes of the $\bP^{1}$ factors making up the ambient space $\cA$. Because the line bundle associated to the constraint for the hypersurface $\cM$ is ample, the Lefschetz hyperplane theorem  \cite{Hubsch:1992nu,bott1959}, can be employed to establish an equality between the dimensions of the effective, or Mori, cones of $\cA$ and $\cM$.
However, as in the example reported by Ref.~\cite{Anderson:2015iia}, upon descent from $\cA$ to $\cM$, although the \emph{dimension} of the Mori cone stays the same, the \emph{width} of the cone increases.
The increase in the width is such to accommodate line bundles with negative charges, which are necessary to build the gCICY.
By analyzing the Koszul sequence and by using the modified pyCICY package \cite{CICYtoolkit} it was found that line bundles $\cO(a_1,a_2,a_3,b,c)$ satisfying:
\begin{multline}
\left\{a_i\ge0,\ b\ge0,\ c\ge0\right\}\cup\left(\bigcup_{i=1}^3\left\{a_i\leq -1,\ a_j\ge-a_i,\ b\ge0, c\ge -3a_i,\ j\neq i\right\}\right)\\\cup \left\{c\leq -1,\ a_i\ge -c,\ i=1,\dots,3,\ b\ge 0\right\}
\end{multline}
all have a positive $h^0$ on $\cM$, and are thus part of the Mori cone.

To aid in the search of the Mori cone, an intuitive pattern which was noted between the zeroth Hodge numbers of line bundles with negative charges was also used.
A formal proof of this pattern is not to be offered in this paper, but, in all cases tested, this pattern allowed for the prediction of a lower bound for the zeroth Hodge number of line bundles on hypersurfaces, which was always saturated.
Consider the line bundle $\cO_\cM\left( 1,1,1,0,-1 \right)$, for which $h^{0}\left(\cM, \cO\left( 1,1,1,0,-1 \right) \right)=3$.
A tentative rational section of $\cO_\cM\left( 1,1,1,1,-1 \right)$ could be built by multiplication of a section of $\cO\left( 1,1,1,0,-1 \right)$ with one of $\cO_\cM\left( 0,0,0,1,0 \right)$.
Given that $h^{0}\left( \cM, \cO\left(0,0,0,1,0  \right) \right)=2$, six sections can be constructed in such manner, establishing a lower bound for $h^{0}\left( \cM,\cO\left( 1,1,1,1,-1 \right) \right)\ge6$. In this example, this lower bound can in fact be shown to be saturated: as calculated by the pyCICY toolkit, $h^{0}\left( \cM,\cO\left( 1,1,1,1,-1 \right) \right)=6$. 

Notwithstanding the increase in the width of the Mori cone, it can be shown using the constraints \eqref{eq:posconstr} that the K{\"a}hler cone for the hypersurface $\cM$ remains the set of line bundles with positive charges.
In particular, line bundles with negative charges can be excluded by finding curves onto which \eqref{eq:posconstr} fails.
Consider, for example, the line bundle $\cO_\cM\left( 1,1,1,1,-1 \right)$.
To show that $\cO_\cM\left( 1,1,1,1,-1 \right)$  is not in the K{\"a}hler cone, let $J=J_1+J_2+J_3+J_4-J_5$, where the $J_{i}$ are the curvature two forms for $\bP^{1}$, and let $C\subset \cM$ be the curve defined by: 
\begin{equation}
    C=\cM\cap\left\{p_1=0\right\}\cap\left\{p_2=0\right\}\cap\left\{p_3=0\right\}
\end{equation}
where $p_1,p_2,p_3$ are sections of $\cO_\cM(1,1,1,0,-1)$\footnote{Note that, because $h^0(\cM,\cO(1,1,1,0,-1))=3$, $C$ can be chosen to be irreducible.}.
Integration of $J$ over $C$ gives:
\begin{equation}
  \int_C J=\int_\cM \left( J_1+J_2+J_3+J_4-J_5 \right)\wedge\left( J_1+J_2+J_3-J_5 \right)^3=0.
\end{equation}
Therefore, by the positivity constraints \eqref{eq:posconstr} we conclude that although $\cO_\cM\left( 1,1,1,1,-1 \right)$ appears in the Mori cone of $\cM$ it does not appear in $\cM$'s K{\"a}hler cone.
In a similar way, one can exclude all other line bundles with negative charges from the K{\"a}hler cone.

Line bundles with positive charges can be shown to be part of the K{\"a}hler cone using equations \eqref{eq:posconstr}.
While the positivity of integrals against submanifolds obtained from line bundles with positive charges is trivial, positivity had to be checked for integration against submanifolds obtained form line bundles with negative charges.
This was done by expressing the line bundles in the Mori cone in terms of variable charges, and studying the polynomials in these variables obtained from integration.
As an example, let $J=\sum_{i=1}^5J_i$ and let $\cN$ be a  hypersurface in $\cM$ obtained from the zero set of a section of $\mathcal{O}_\cM(x+a_1,x+a_2,x+a_3,a_4,-x)$, for $x\in \mathbb{N}^*$ and $a_i\in\mathbb{N}$.
Then, within the bounds specified, the integral:
\begin{equation*}
    \int_\cN J\wedge J\wedge J=6 \left(5 a_1+5 a_2+5 a_3+6 a_4+12 x\right)
\end{equation*}
is always positive.
Repeating this calculation with a sufficient number of submanifolds to cover all of the negative line bundles in the Mori cone, one can establish that $J=\sum_{i=1}^5J_i$ is part of the K{\"a}hler cone. Generalising to $J=\sum_{i=1}^5 c_i J_i$, with positive $c_i$ to cover the full positive orthant, is straightforward.

Having found the K{\"a}hler and Mori cones of $\cM$ the analysis can be completed by repeating these steps on the full gCICY $X_2$, considering the polynomial constraint as well as the rational one.
However, due to technical limitations in the construction of sections with arbitrary denominators as well as computational limitations in the calculation of Hodge numbers for line bundles with high charges, we were only able to establish an approximation for the Mori cone and the K\"ahler cone. To illustrate the problem encountered, we present a specific check of equations \eqref{eq:posconstr}. 

Take the line bundles $\cO_{X_2}\left(-1,1,1,0,3\right)$ and $\cO_{X_2}\left(-1,3,3,-1,5\right)$, that admit $h^0=1$ and $h^0 = 2$ rational sections, respectively. Denote these rational section $r$ and $s_1, s_2$. We can now test equations \ref{eq:posconstr} by considering the curve 
\begin{align} \label{eq:C}
    C = X_2 \cap \{ r = 0 \} \cap \{ s_{1,2} = 0 \} .
\end{align}
Given such a curve we can compute the integral for any two-form $J$ related to a line bundle $\cO (a_1, a_2, a_3, a_4, a_5)$, to find
\begin{align}
\label{eq:nint}
  \int_C J = 82 a_{1}-18 a_{2}-18 a_{3}+36 a_{4}-6 a_{5}.
\end{align}
Therefore, there are certain positive values of the charges which lead to a negative integral, and thus invalidate the last of equations \eqref{eq:posconstr}. We have been able to identify two possible explanations for the negative integral:
it could either manifest a restriction on the K\"ahler cone, or some non trivial dependencies arising among curves in the gCICY.

In the first case, the negative integral arising in \eqref{eq:nint} establishes a restriction of the Kähler cone from the positive orthant to the subset given by
\begin{equation} \label{eq:kahconst}
  82a_{1}+36a_{4}>18(a_{2}+a_{3})+6a_{5}.
\end{equation}
If confirmed, this constraint would then have to be checked when establishing the stability of vector bundles.

However, an alternative explanation, which would invalidate the constraint \eqref{eq:kahconst},  could  be that $C$ is not an irreducible surface.
Even though in this particular case $\cO_{X_2}\left( -1,3,3,-1,5 \right)$ has two independent sections, generating two independent ideals, these could still be contained in the ideal generated by the rational constraint defining $r$.\footnote{In a similar example consider $\mathbb{F}\left[ x,y \right]$, $I_{1}=\left<xy^{2}\right>$, $I_{2}=\left<x^{2}y\right>$ and $I_{3}=\left<xy\right>$ then $I_{1}$ and $I_{2}$ are independent, but both of them are contained in $I_{3}$.} In this case, \eqref{eq:C} would not define a surface, but a four-dimensional submanifold of $X$. Thus, it is important to study these ideals to make sure that they are independent. To accomplish this we must construct the rational sections for the line bundles defining $C$, using the methods described in section \ref{sec:twogcicys}.

The construction of rational sections is subtle. Consider the case of $\cO_{X_2}\left( -1,1,1,0,3 \right)$, for which the zeroth Hodge number is one.
Even though we expect a single independent section $r$ on the gCICY, three sections $r_{i}$ can be built using the homogeneous coordinates of the ambient space.
To build these, we start by selecting a set of denominators $\Delta'=\left\{ x_{0,0},x_{0,1}-x_{0,0},x_{0,1}+x_{0,0} \right\}$ and by dividing the polynomial constraint to find $m'_{i}$ and $N'_{i}$, defined by:
\begin{align}
  p=m'_{i}\Delta_{i}^{'}+N'_{i},\ \Delta_{i}'\in\Delta'.
\end{align}
These functions are used to build the numerators for $r_{i}$.
Using the fact that the denominators in $\Delta'$ cannot vanish together on a complex projective space, we can divide out from the $N'_{i}$ terms any further dependence on the zeroth complex projective space and thus obtaining sections of $\cO_{X_2}\left( 0,1,1,0,3 \right)$ to be used as the numerators of $r_{i}$.

This construction produces three seemingly independent sections, however, one can again use techniques from computational algebra and show that on $X_2$ all three sections are equivalent.
This technique verifies the independence of sections irrespective of the fact that they belong to the same line bundle, and therefore we will go through it in detail.
After having explicitly constructed $p$, the polynomial constraint, $q_{i}$, the independent rational sections of $\cO\left( 1,1,1,2,-1 \right)$ and the three $r_{j}$s, the ambient space can be divided into 16 regions $R_{ij},\ i,j=0,\dots,3$ corresponding to the vanishing of the denominators of $r_{j}$ and the sections $q_{i}$ in a similar way as in \eqref{eq:regiondivision}.
In each region, an ideal $I_{i,j,U_{l}}$ can be built using from the set $\left\{p,q_{i},r_{j},\rho_{ij},\phi_{U_{l}} \right\}$, where $\rho_{ij}$ localizes the computation to the region $R_{ij}$  and $\phi_{U_{l}}$, as in \ref{sec:twogcicys}, localizes the computation further into a particular open set $U_{l}$ of an open cover of the ambient space.
For each region $R_{ij}$, each open set in the cover $U_{l}$, and each constraint $r_{i}$, one can check that $r_{k}\in I_{i,j,U_{l}},\ k=0,\dots,3$.
This implies that the vanishing of a rational section $r_{i}$ implies the vanishing of all of the others, and therefore they are not independent.
Consequently, this calculation further provides a non trivial consistency check that $h^0(\cO_{X_2}\left(-1,1,1,0,3\right)) = 1$.

In similar fashion one can attempt to construct the two sections of $\cO\left( -1,3,3,-1,5 \right)$.
However, here, one runs into additional issues.
$\cO_{X_2}(-1,3,3,-1,5)$ has a negative degree in the coordinates of $\mathbb{P}^1_4$ which do not contribute to $p$.
We must then use the rational sections $q_\alpha$ to define the (double) rational sections $s_{1,2}$. Proceeding as above then leads to a more severe overcounting of ``naive'' rational sections, deriving from the three $q_{\alpha}$ and the additional increase in the homogeneous coordinates needed to cancel the contribution of the denominators present in the rational sections $q_{\alpha}$ and to obtain the correct degrees. 
Furthermore, one now has an additional cubic increase in regions $R_i$ to consider due to the three\footnote{By using $q_{\alpha}$, the coordinates of $\mathbb{P}^1_1 , \mathbb{P}^1_4, \mathbb{P}^1_5$ will all appear in the denominator.} fractions with $\mathbb{P}^1$ coordinates in the denominator.
In a last step one then has to confirm that also $r$ and $s_{1,2}$ are independent on $X_2$. 

Unfortunately, even in the finite range of line bundles we can computationally control, there are several additional curves for which (semi-)positive divisors fail to lead to positive intersection numbers in a similar manner. Determining whether these restrict the width of the Kähler cone then quickly becomes intractable, barring  a systematic method to construct rational sections. 
Hence, we decided to only check the necessary condition given in equation \ref{eq:slope} in the scans of this paper.

\section{Heterotic Line Bundle Models}

\begin{longtable}{| p{.05\textwidth} | p{.90\textwidth} |} 
    \centering
    KILLED & LINE!!!! \kill
    \caption[]{All heterotic standard-like line bundle models on the gCICY $X_1$ with freely acting symmetry $\mathbb{Z}_2$.}
    \label{tab:models1}
    \\ \hline
    $V_i$ & $(L_1, L_2, L_3, L_4, L_5)$ 
    \endfirsthead
        \hline
    $V_i$ & $(L_1, L_2, L_3, L_4, L_5)$ \\
    \hline \hline
    \endhead
    \hline
$V_{1}$ & ((-3, 0, 1, 1, -1), (-1, 1, -2, 0, 1), (1, -1, 0, 0, 1), (1, 0, 0, -1, 1), (2, 0, 1, 0, -2))\\
$V_{2}$ & ((-3, 0, 1, 1, -1), (0, 0, -1, 1, 1), (0, 1, -1, -2, 1), (1, -1, 0, 0, 1), (2, 0, 1, 0, -2))\\
$V_{3}$ & ((-2, -1, 0, 1, 1), (0, 1, -1, 0, 1), (0, 1, 0, -1, 1), (1, -3, 1, 0, -1), (1, 2, 0, 0, -2))\\
$V_{4}$ & ((-2, -1, 0, 2, 1), (-1, 1, 0, 0, -1), (1, 0, -2, 0, 2), (1, 0, 1, -1, -1), (1, 0, 1, -1, -1))\\
$V_{5}$ & ((-2, -1, 0, 2, 1), (0, -1, 0, 0, 1), (0, -1, 0, 0, 1), (0, 1, 1, -2, -1), (2, 2, -1, 0, -2))\\
$V_{6}$ & ((-2, -1, 0, 2, 1), (0, -1, 0, 0, 1), (0, -1, 0, 0, 1), (1, 1, 1, -1, -3), (1, 2, -1, -1, 0))\\
$V_{7}$ & ((-2, -1, 1, 1, 0), (-1, 1, 1, -1, 0), (1, -1, -1, 0, 1), (1, -1, -1, 0, 1), (1, 2, 0, 0, -2))\\
$V_{8}$ & ((-2, -1, 1, 1, 0), (0, -1, 0, 0, 1), (0, -1, 0, 0, 1), (0, 1, 0, -1, 0), (2, 2, -1, 0, -2))\\
$V_{9}$ & ((-2, -1, 1, 2, 0), (0, -1, 1, 0, 0), (0, -1, 1, 0, 0), (0, 1, -1, -2, 1), (2, 2, -2, 0, -1))\\
$V_{10}$ & ((-2, -1, 1, 2, 0), (0, -1, 1, 0, 0), (0, -1, 1, 0, 0), (1, 1, -3, -1, 1), (1, 2, 0, -1, -1))\\
$V_{11}$ & ((-2, 0, 0, 2, 1), (0, -1, 2, 1, -2), (0, 1, 0, -1, -1), (1, 0, -1, -1, 1), (1, 0, -1, -1, 1))\\
$V_{12}$ & ((-2, 0, 1, 1, -1), (-1, 0, 0, 1, 0), (0, 0, -1, 1, 1), (1, 1, 0, -1, -2), (2, -1, 0, -2, 2))\\
$V_{13}$ & ((-2, 0, 1, 1, -1), (0, 0, 0, -1, 1), (0, 0, 0, -1, 1), (0, 0, 0, -1, 1), (2, 0, -1, 2, -2))\\
$V_{14}$ & ((-2, 1, 0, 1, -1), (-1, -1, 0, 1, 1), (1, 0, -2, 0, 2), (1, 0, 1, -1, -1), (1, 0, 1, -1, -1))\\
$V_{15}$ & ((-2, 1, 1, -1, 0), (-1, -1, 1, 1, 0), (1, 0, -1, -1, 1), (1, 0, -1, -1, 1), (1, 0, 0, 2, -2))\\
$V_{16}$ & ((-2, 1, 1, 1, -2), (0, -1, 1, 0, 1), (0, 0, 1, -1, 1), (1, 0, -2, 0, 2), (1, 0, -1, 0, -2))\\
$V_{17}$ & ((-1, -2, 0, 1, 2), (-1, 0, 0, 1, 0), (-1, 0, 0, 1, 0), (1, 1, 1, -3, -1), (2, 1, -1, 0, -1))\\
$V_{18}$ & ((-1, -2, 2, 0, 1), (-1, 0, 0, 0, 1), (-1, 0, 0, 0, 1), (1, 1, -1, 1, -3), (2, 1, -1, -1, 0))\\
$V_{19}$ & ((-1, -2, 2, 1, 0), (-1, 0, 0, 1, 0), (-1, 0, 0, 1, 0), (1, 1, -1, -3, 1), (2, 1, -1, 0, -1))\\
$V_{20}$ & ((-1, -1, 0, 0, 1), (-1, 1, 0, 0, 0), (0, -1, 1, 0, 0), (0, -1, 1, 0, 0), (2, 2, -2, 0, -1))\\
$V_{21}$ & ((-1, -1, 0, 0, 1), (-1, 1, 0, 1, -1), (0, -1, 1, 0, 0), (0, -1, 1, 0, 0), (2, 2, -2, -1, 0))\\
$V_{22}$ & ((-1, -1, 0, 0, 1), (-1, 1, 1, 0, -1), (0, -1, 0, 1, 0), (0, -1, 0, 1, 0), (2, 2, -1, -2, 0))\\
$V_{23}$ & ((-1, -1, 0, 0, 1), (-1, 2, -1, 0, 1), (0, -1, 0, 1, 0), (1, -1, 0, 2, -1), (1, 1, 1, -3, -1))\\
$V_{24}$ & ((-1, -1, 0, 0, 1), (0, -1, 0, 1, 0), (0, -1, 0, 1, 0), (0, 2, -1, 1, 0), (1, 1, 1, -3, -1))\\
$V_{25}$ & ((-1, -1, 0, 1, 0), (-1, 0, 0, 0, 1), (-1, 0, 0, 0, 1), (1, 1, 1, -1, -3), (2, 0, -1, 0, 1))\\
$V_{26}$ & ((-1, -1, 0, 1, 0), (-1, 0, 0, 0, 1), (-1, 1, 0, -1, 2), (1, 1, 1, -1, -3), (2, -1, -1, 1, 0))\\
$V_{27}$ & ((-1, -1, 0, 1, 0), (-1, 0, 1, 0, 0), (-1, 0, 1, 0, 0), (1, -1, 0, 0, 0), (2, 2, -2, -1, 0))\\
$V_{28}$ & ((-1, -1, 0, 1, 1), (-1, -1, 0, 1, 1), (-1, 0, 1, 0, -1), (1, 2, -1, 0, -2), (2, 0, 0, -2, 1))\\
$V_{29}$ & ((-1, -1, 0, 1, 1), (-1, -1, 0, 1, 1), (0, 2, 0, -2, 1), (1, -1, 1, 0, -2), (1, 1, -1, 0, -1))\\
$V_{30}$ & ((-1, -1, 0, 1, 1), (-1, -1, 0, 1, 1), (0, 2, 0, 1, -2), (1, -1, 1, -2, 0), (1, 1, -1, -1, 0))\\
$V_{31}$ & ((-1, -1, 0, 2, 1), (-1, 0, 0, 1, 0), (-1, 1, 0, 0, -1), (1, -1, 1, -3, 1), (2, 1, -1, 0, -1))\\
$V_{32}$ & ((-1, -1, 0, 2, 1), (0, 0, -1, 0, 1), (0, 0, -1, 0, 1), (0, 0, 1, -1, 0), (1, 1, 1, -1, -3))\\
$V_{33}$ & ((-1, -1, 1, 0, 0), (-1, 0, 0, 1, 0), (-1, 0, 0, 1, 0), (1, -1, 0, 0, 0), (2, 2, -1, -2, 0))\\
$V_{34}$ & ((-1, -1, 1, 1, 0), (-1, -1, 1, 1, 0), (0, -1, 0, -1, 1), (0, 2, -2, 1, 0), (2, 1, 0, -2, -1))\\
$V_{35}$ & ((-1, -1, 1, 2, 0), (0, -1, 1, -1, 0), (0, 0, 0, -1, 1), (0, 1, -1, -1, 2), (1, 1, -1, 1, -3))\\
$V_{36}$ & ((-1, -1, 1, 2, 0), (0, 0, 0, -1, 1), (0, 0, 0, -1, 1), (0, 0, 0, -1, 1), (1, 1, -1, 1, -3))\\
$V_{37}$ & ((-1, 0, -1, 0, 1), (-1, 0, 0, 1, 0), (-1, 0, 0, 1, 0), (1, 0, -1, 0, 0), (2, 0, 2, -2, -1))\\
$V_{38}$ & ((-1, 0, -1, 1, 0), (-1, 0, 0, 0, 1), (-1, 0, 0, 0, 1), (1, 0, -1, 0, 0), (2, 0, 2, -1, -2))\\
$V_{39}$ & ((-1, 0, -1, 1, 0), (-1, 0, 0, 0, 1), (-1, 0, 0, 0, 1), (1, 1, -1, -1, 0), (2, -1, 2, 0, -2))\\
$V_{40}$ & ((-1, 0, -1, 1, 0), (-1, 0, 0, 0, 1), (-1, 0, 0, 0, 1), (1, 1, 1, -1, -3), (2, -1, 0, 0, 1))\\
$V_{41}$ & ((-1, 0, 0, 0, 1), (-1, 0, 0, 0, 1), (-1, 0, 0, 0, 1), (0, 0, 0, 1, -4), (3, 0, 0, -1, 1))\\
$V_{42}$ & ((-1, 0, 0, 0, 1), (-1, 0, 0, 0, 1), (-1, 0, 0, 0, 1), (0, 0, 1, 0, -4), (3, 0, -1, 0, 1))\\
$V_{43}$ & ((-1, 0, 0, 0, 1), (-1, 0, 0, 0, 1), (-1, 0, 0, 0, 1), (0, 1, 0, 0, -4), (3, -1, 0, 0, 1))\\
$V_{44}$ & ((-1, 0, 0, 0, 1), (-1, 0, 0, 0, 1), (-1, 0, 0, 0, 1), (1, -1, 1, 1, -3), (2, 1, -1, -1, 0))\\
$V_{45}$ & ((-1, 0, 0, 0, 1), (-1, 0, 0, 0, 1), (-1, 0, 0, 0, 1), (1, 0, -2, 1, -1), (2, 0, 2, -1, -2))\\
$V_{46}$ & ((-1, 0, 0, 0, 1), (-1, 0, 0, 0, 1), (-1, 0, 0, 0, 1), (1, 1, -2, 0, -1), (2, -1, 2, 0, -2))\\
$V_{47}$ & ((-1, 0, 0, 0, 1), (-1, 0, 0, 0, 1), (-1, 0, 0, 0, 1), (1, 1, -1, 1, -3), (2, -1, 1, -1, 0))\\
$V_{48}$ & ((-1, 0, 0, 0, 1), (-1, 0, 0, 0, 1), (-1, 0, 0, 0, 1), (1, 1, 1, -1, -3), (2, -1, -1, 1, 0))\\
$V_{49}$ & ((-1, 0, 0, 0, 1), (-1, 0, 0, 0, 1), (-1, 0, 1, -1, 1), (0, 0, 0, 1, -4), (3, 0, -1, 0, 1))\\
$V_{50}$ & ((-1, 0, 0, 0, 1), (-1, 0, 0, 0, 1), (-1, 1, -2, 1, 0), (1, -1, 0, 0, 0), (2, 0, 2, -1, -2))\\
$V_{51}$ & ((-1, 0, 0, 0, 1), (-1, 0, 0, 0, 1), (-1, 1, -2, 1, 0), (1, 0, 0, -1, 0), (2, -1, 2, 0, -2))\\
$V_{52}$ & ((-1, 0, 0, 0, 1), (-1, 0, 0, 0, 1), (-1, 1, -1, 0, 0), (1, -1, 1, 1, -3), (2, 0, 0, -1, 1))\\
$V_{53}$ & ((-1, 0, 0, 0, 1), (-1, 0, 0, 0, 1), (-1, 1, -1, 0, 0), (1, 0, -1, 0, 0), (2, -1, 2, 0, -2))\\
$V_{54}$ & ((-1, 0, 0, 0, 1), (-1, 0, 0, 0, 1), (-1, 1, -1, 0, 1), (0, 0, 1, 0, -4), (3, -1, 0, 0, 1))\\
$V_{55}$ & ((-1, 0, 0, 0, 1), (-1, 0, 0, 0, 1), (-1, 1, 0, -1, 1), (0, 0, 0, 1, -4), (3, -1, 0, 0, 1))\\
$V_{56}$ & ((-1, 0, 0, 0, 1), (-1, 0, 0, 0, 1), (-1, 1, 0, -1, 1), (1, 0, -2, 1, -1), (2, -1, 2, 0, -2))\\
$V_{57}$ & ((-1, 0, 0, 0, 1), (-1, 0, 0, 0, 1), (-1, 2, -2, 0, 1), (1, -1, 1, 1, -3), (2, -1, 1, -1, 0))\\
$V_{58}$ & ((-1, 0, 0, 0, 1), (-1, 0, 0, 0, 1), (0, -1, -1, 2, 1), (1, 0, 0, -1, 0), (1, 1, 1, -1, -3))\\
$V_{59}$ & ((-1, 0, 0, 0, 1), (-1, 0, 0, 0, 1), (0, -1, 2, -1, 1), (1, 0, -1, 0, 0), (1, 1, -1, 1, -3))\\
$V_{60}$ & ((-1, 0, 0, 0, 1), (-1, 0, 0, 0, 1), (0, 2, -1, -1, 1), (1, -1, 0, 0, 0), (1, -1, 1, 1, -3))\\
$V_{61}$ & ((-1, 0, 0, 1, -1), (-1, 2, 1, -1, 0), (0, 1, -1, 0, 1), (1, -2, 0, 0, 2), (1, -1, 0, 0, -2))\\
$V_{62}$ & ((-1, 0, 0, 1, 0), (-1, 0, 0, 1, 0), (-1, 0, 0, 1, 0), (0, 0, 0, -4, 1), (3, 0, 0, 1, -1))\\
$V_{63}$ & ((-1, 0, 0, 1, 0), (-1, 0, 0, 1, 0), (-1, 0, 0, 1, 0), (0, 0, 1, -4, 0), (3, 0, -1, 1, 0))\\
$V_{64}$ & ((-1, 0, 0, 1, 0), (-1, 0, 0, 1, 0), (-1, 0, 0, 1, 0), (0, 1, 0, -4, 0), (3, -1, 0, 1, 0))\\
$V_{65}$ & ((-1, 0, 0, 1, 0), (-1, 0, 0, 1, 0), (-1, 0, 0, 1, 0), (1, -2, 1, -1, 0), (2, 2, -1, -2, 0))\\
$V_{66}$ & ((-1, 0, 0, 1, 0), (-1, 0, 0, 1, 0), (-1, 0, 0, 1, 0), (1, -1, 1, -3, 1), (2, 1, -1, 0, -1))\\
$V_{67}$ & ((-1, 0, 0, 1, 0), (-1, 0, 0, 1, 0), (-1, 0, 0, 1, 0), (1, 0, -2, -1, 1), (2, 0, 2, -2, -1))\\
$V_{68}$ & ((-1, 0, 0, 1, 0), (-1, 0, 0, 1, 0), (-1, 0, 0, 1, 0), (1, 1, -1, -3, 1), (2, -1, 1, 0, -1))\\
$V_{69}$ & ((-1, 0, 0, 1, 0), (-1, 0, 0, 1, 0), (-1, 0, 0, 1, 0), (1, 1, 0, -1, -2), (2, -1, 0, -2, 2))\\
$V_{70}$ & ((-1, 0, 0, 1, 0), (-1, 0, 0, 1, 0), (-1, 0, 0, 1, 0), (1, 1, 1, -3, -1), (2, -1, -1, 0, 1))\\
$V_{71}$ & ((-1, 0, 0, 1, 0), (-1, 0, 0, 1, 0), (-1, 1, 0, 0, -1), (1, -1, 1, -3, 1), (2, 0, -1, 1, 0))\\
$V_{72}$ & ((-1, 0, 0, 1, 0), (-1, 0, 0, 1, 0), (0, -1, 2, 1, -1), (1, 0, -1, 0, 0), (1, 1, -1, -3, 1))\\
$V_{73}$ & ((-1, 0, 0, 1, 0), (-1, 0, 0, 1, 0), (0, 0, -1, 1, 1), (0, 1, 1, -1, -3), (2, -1, 0, -2, 2))\\
$V_{74}$ & ((-1, 0, 0, 1, 0), (-1, 0, 0, 1, 0), (0, 0, -1, 1, 1), (1, -1, 0, 0, 0), (1, 1, 1, -3, -1))\\
$V_{75}$ & ((-1, 0, 0, 1, 0), (-1, 0, 0, 1, 0), (0, 1, 0, 1, -1), (1, -1, 1, -3, 1), (1, 0, -1, 0, 0))\\
$V_{76}$ & ((-1, 0, 0, 1, 0), (-1, 0, 0, 1, 0), (0, 2, -1, 1, -1), (1, -1, 0, 0, 0), (1, -1, 1, -3, 1))\\
$V_{77}$ & ((-1, 0, 0, 1, 0), (-1, 0, 1, -1, 0), (0, 0, 0, -1, 1), (0, 0, 0, -1, 1), (2, 0, -1, 2, -2))\\
$V_{78}$ & ((-1, 0, 0, 1, 0), (0, 0, 0, -1, 1), (0, 0, 0, -1, 1), (0, 1, -1, 0, 1), (1, -1, 1, 1, -3))\\
$V_{79}$ & ((-1, 0, 0, 1, 1), (-1, 1, 1, 1, -3), (0, -1, 0, 0, 1), (1, -2, 0, -1, 1), (1, 2, -1, -1, 0))\\
$V_{80}$ & ((-1, 0, 0, 1, 1), (0, -1, 1, 0, 0), (0, 0, -1, 0, 1), (0, 0, -1, 0, 1), (1, 1, 1, -1, -3))\\
$V_{81}$ & ((-1, 0, 0, 1, 1), (0, -1, 1, 0, 0), (0, 0, -1, 1, 0), (0, 0, -1, 1, 0), (1, 1, 1, -3, -1))\\
$V_{82}$ & ((-1, 0, 0, 2, 1), (0, -1, 1, -1, 0), (0, 0, 0, -1, 1), (0, 0, 0, -1, 1), (1, 1, -1, 1, -3))\\
$V_{83}$ & ((-1, 0, 0, 2, 1), (0, 0, 0, -1, 1), (0, 0, 0, -1, 1), (0, 1, -1, -1, 0), (1, -1, 1, 1, -3))\\
$V_{84}$ & ((-1, 0, 0, 3, 1), (0, 0, 0, -1, 1), (0, 0, 0, -1, 1), (0, 0, 0, -1, 1), (1, 0, 0, 0, -4))\\
$V_{85}$ & ((-1, 0, 1, 1, 0), (0, 0, 1, 1, -1), (0, 1, -2, 2, 0), (0, 1, -1, -3, 1), (1, -2, 1, -1, 0))\\
$V_{86}$ & ((-1, 0, 2, 2, -2), (0, -1, 0, 1, 0), (0, 0, 0, -1, 1), (0, 0, 0, -1, 1), (1, 1, -2, -1, 0))\\
$V_{87}$ & ((-1, 0, 2, 2, -2), (0, -1, 1, -1, 2), (0, 0, 0, -1, 1), (0, 1, -1, -1, 0), (1, 0, -2, 1, -1))\\
$V_{88}$ & ((-1, 0, 2, 2, -2), (0, 0, -1, 1, 0), (0, 0, 0, -1, 1), (0, 0, 0, -1, 1), (1, 0, -1, -1, 0))\\
$V_{89}$ & ((-1, 0, 2, 2, -2), (0, 0, 0, -1, 1), (0, 0, 0, -1, 1), (0, 0, 0, -1, 1), (1, 0, -2, 1, -1))\\
$V_{90}$ & ((-1, 0, 2, 2, -2), (0, 0, 0, -1, 1), (0, 0, 0, -1, 1), (0, 1, -1, -1, 0), (1, -1, -1, 1, 0))\\
$V_{91}$ & ((-1, 1, -1, 2, 0), (-1, 1, 1, 1, -3), (0, 0, 0, -1, 1), (0, 0, 0, -1, 1), (2, -2, 0, -1, 1))\\
$V_{92}$ & ((-1, 1, 0, -1, 2), (-1, 1, 1, 1, -3), (0, -1, 0, 1, -1), (0, 1, -1, 0, 1), (2, -2, 0, -1, 1))\\
$V_{93}$ & ((-1, 1, 1, 1, -3), (-1, 2, 1, -1, 0), (0, -1, 0, 0, 1), (0, -1, 0, 0, 1), (2, -1, -2, 0, 1))\\
$V_{94}$ & ((-1, 1, 1, 1, -3), (0, -1, 0, 0, 1), (0, -1, 0, 0, 1), (0, -1, 0, 0, 1), (1, 2, -1, -1, 0))\\
$V_{95}$ & ((-1, 1, 1, 1, -3), (0, -1, 0, 0, 1), (0, -1, 0, 0, 1), (0, 1, -1, 0, 0), (1, 0, 0, -1, 1))\\
$V_{96}$ & ((-1, 1, 1, 1, -3), (0, -1, 0, 0, 1), (0, -1, 0, 0, 1), (0, 2, 0, -1, 1), (1, -1, -1, 0, 0))\\
$V_{97}$ & ((-1, 1, 1, 1, -3), (0, -1, 0, 2, 1), (0, 0, 0, -1, 1), (0, 0, 0, -1, 1), (1, 0, -1, -1, 0))\\
$V_{98}$ & ((-1, 1, 1, 1, -3), (0, 0, 0, -1, 1), (0, 0, 0, -1, 1), (0, 0, 0, -1, 1), (1, -1, -1, 2, 0))\\
$V_{99}$ & ((0, -1, 0, 0, 1), (0, -1, 0, 0, 1), (0, -1, 0, 0, 1), (0, 0, 0, 1, -4), (0, 3, 0, -1, 1))\\
\hline
\end{longtable}

\newpage

\begin{longtable}{| p{.05\textwidth} | p{.90\textwidth} |} 
    \centering
    KILLED & LINE!!!! \kill
    \caption[]{All heterotic standard-like line bundle models on the gCICY $X_2$ with freely acting symmetry $\mathbb{Z}_2$.}
    \label{tab:models2}
    \\ \hline
    $V_i$ & $(L_1, L_2, L_3, L_4, L_5)$ 
    \endfirsthead
        \hline
    $V_i$ & $(L_1, L_2, L_3, L_4, L_5)$ \\
    \hline \hline
    \endhead
    \hline
$V_{1}$ & ((-3, 1, 1, -1, -1), (0, 2, -1, 1, 1), (1, -1, 0, 0, 0), (1, -1, 0, 0, 0), (1, -1, 0, 0, 0))\\
$V_{2}$ & ((-2, 0, 1, -1, 1), (0, 0, -1, 1, 0), (0, 0, -1, 1, 0), (0, 0, -1, 1, 0), (2, 0, 2, -2, -1))\\
$V_{3}$ & ((-2, 0, 1, 1, -1), (0, 0, -1, 0, 1), (0, 0, -1, 0, 1), (0, 0, -1, 0, 1), (2, 0, 2, -1, -2))\\
$V_{4}$ & ((-1, -1, 0, 1, 2), (0, -1, 1, 0, 0), (0, -1, 1, 0, 0), (0, 2, 1, 0, -1), (1, 1, -3, -1, -1))\\
$V_{5}$ & ((-1, -1, 0, 2, 1), (-1, 0, 1, 0, 0), (-1, 0, 1, 0, 0), (1, 1, -3, -1, -1), (2, 0, 1, -1, 0))\\
$V_{6}$ & ((-1, -1, 0, 2, 1), (0, -1, 1, 0, 0), (0, -1, 1, 0, 0), (0, 2, 1, -1, 0), (1, 1, -3, -1, -1))\\
$V_{7}$ & ((-1, -1, 1, 1, 2), (0, 0, -1, 0, 1), (0, 0, -1, 0, 1), (0, 1, 0, 0, -4), (1, 0, 1, -1, 0))\\
$V_{8}$ & ((-1, 0, -1, 0, 1), (-1, 0, 1, 0, 0), (0, 0, -1, 1, 0), (0, 0, -1, 1, 0), (2, 0, 2, -2, -1))\\
$V_{9}$ & ((-1, 0, -1, 1, 0), (-1, 0, 1, 0, 0), (0, 0, -1, 0, 1), (0, 0, -1, 0, 1), (2, 0, 2, -1, -2))\\
$V_{10}$ & ((-1, 0, 0, 0, 1), (-1, 0, 0, 0, 1), (-1, 0, 0, 0, 1), (0, 0, 0, 1, -4), (3, 0, 0, -1, 1))\\
$V_{11}$ & ((-1, 0, 0, 0, 1), (-1, 0, 0, 0, 1), (-1, 0, 0, 0, 1), (1, -1, 1, 1, -1), (2, 1, -1, -1, -2))\\
$V_{12}$ & ((-1, 0, 0, 0, 1), (-1, 0, 0, 0, 1), (-1, 0, 0, 0, 1), (1, 0, -2, 1, -1), (2, 0, 2, -1, -2))\\
$V_{13}$ & ((-1, 0, 0, 0, 1), (-1, 0, 0, 0, 1), (-1, 0, 0, 0, 1), (1, 1, -1, 1, -1), (2, -1, 1, -1, -2))\\
$V_{14}$ & ((-1, 0, 0, 0, 1), (-1, 0, 0, 0, 1), (0, 0, 1, -2, -1), (1, -1, 0, 1, 0), (1, 1, -1, 1, -1))\\
$V_{15}$ & ((-1, 0, 0, 0, 1), (-1, 0, 0, 0, 1), (0, 0, 1, -1, -1), (1, -1, 0, 0, 0), (1, 1, -1, 1, -1))\\
$V_{16}$ & ((-1, 0, 0, 1, 0), (-1, 0, 0, 1, 0), (-1, 0, 0, 1, 0), (0, 0, 0, -4, 1), (3, 0, 0, 1, -1))\\
$V_{17}$ & ((-1, 0, 0, 1, 0), (-1, 0, 0, 1, 0), (-1, 0, 0, 1, 0), (1, 0, -2, -1, 1), (2, 0, 2, -2, -1))\\
$V_{18}$ & ((-1, 0, 0, 1, 0), (-1, 0, 0, 1, 0), (-1, 0, 0, 1, 0), (1, 1, -1, -1, 1), (2, -1, 1, -2, -1))\\
$V_{19}$ & ((-1, 0, 0, 1, 0), (-1, 0, 0, 1, 0), (0, 0, 0, -4, 1), (0, 1, 0, 1, -1), (2, -1, 0, 1, 0))\\
$V_{20}$ & ((-1, 0, 0, 1, 0), (-1, 0, 0, 1, 0), (0, 0, 1, -1, -1), (1, -1, 0, 0, 0), (1, 1, -1, -1, 1))\\
$V_{21}$ & ((-1, 0, 0, 1, 0), (0, -1, 0, 0, 1), (0, -1, 0, 0, 1), (0, 1, 1, -2, -1), (1, 1, -1, 1, -1))\\
$V_{22}$ & ((-1, 0, 0, 3, 1), (0, 0, 0, -1, 1), (0, 0, 0, -1, 1), (0, 0, 0, -1, 1), (1, 0, 0, 0, -4))\\
$V_{23}$ & ((-1, 0, 1, 0, -1), (-1, 1, 0, 0, -2), (0, 0, 1, -2, 2), (1, -1, -1, 1, 2), (1, 0, -1, 1, -1))\\
$V_{24}$ & ((-1, 0, 1, 0, 0), (-1, 0, 1, 0, 0), (-1, 0, 1, 0, 0), (1, 1, -3, -1, -1), (2, -1, 0, 1, 1))\\
$V_{25}$ & ((-1, 1, 1, -1, 1), (0, -1, 0, 1, 0), (0, -1, 0, 1, 0), (0, -1, 0, 1, 0), (1, 2, -1, -2, -1))\\
$V_{26}$ & ((-1, 1, 1, -1, 1), (0, -1, 1, 0, 0), (0, 0, -1, 1, 0), (0, 0, -1, 1, 0), (1, 0, 0, -1, -1))\\
$V_{27}$ & ((-1, 1, 1, -1, 1), (0, 0, -1, 1, 0), (0, 0, -1, 1, 0), (0, 0, -1, 1, 0), (1, -1, 2, -2, -1))\\
$V_{28}$ & ((-1, 1, 1, 1, -1), (0, -1, 0, 0, 1), (0, -1, 0, 0, 1), (0, -1, 0, 0, 1), (1, 2, -1, -1, -2))\\
$V_{29}$ & ((-1, 1, 2, -2, -1), (0, 0, -1, 1, 0), (0, 0, -1, 1, 0), (0, 0, -1, 1, 0), (1, -1, 1, -1, 1))\\
$V_{30}$ & ((-1, 1, 2, -1, -2), (0, 0, -1, 0, 1), (0, 0, -1, 0, 1), (0, 0, -1, 0, 1), (1, -1, 1, 1, -1))\\
$V_{31}$ & ((0, -2, 1, -1, 1), (0, 0, -1, 1, 0), (0, 0, -1, 1, 0), (0, 0, -1, 1, 0), (0, 2, 2, -2, -1))\\
$V_{32}$ & ((0, -2, 1, 1, -1), (0, 0, -1, 0, 1), (0, 0, -1, 0, 1), (0, 0, -1, 0, 1), (0, 2, 2, -1, -2))\\
$V_{33}$ & ((0, 0, -1, 1, 0), (0, 0, -1, 1, 0), (0, 0, -1, 1, 0), (0, 0, 0, -4, 1), (0, 0, 3, 1, -1))\\
\hline
\end{longtable}

\end{appendix}

\bibliographystyle{ytphys}
\bibliography{ref}

\end{document}